\documentclass[a4paper,11pt]{article}
\pdfoutput=1 

\usepackage{jheppub} 
\usepackage{slashed}

\newcommand{\eexxa}{$e^+e^-\rightarrow \chi\bar{\chi} \gamma\ $}
\newcommand{\eevva}{$e^+e^-\rightarrow\nu\bar{\nu}\gamma\ $}
\newcommand{\eeeea}{$e^+e^-\rightarrow e^+e^-\gamma\ $}

\newcommand{\GeV}{\,{\rm GeV}}

\newcommand{\be}{\begin{equation}}
\newcommand{\ee}{\end{equation}}
\newcommand{\ba}{\begin{array}}
\newcommand{\ea}{\end{array}}
\newcommand{\bea}{\begin{eqnarray}}
\newcommand{\eea}{\end{eqnarray}}

\usepackage{ulem,fancyvrb}
\usepackage{xcolor}

\title{Probing dark matter particles at CEPC}

\author[a,b,c]{Zuowei Liu,}
\author[a]{Yong-Heng Xu} 
\author[d,c]{and Yu Zhang} 

\affiliation[a]{Department of Physics, Nanjing University, Nanjing 210093, China} 
\affiliation[b]{Center for High Energy Physics, Peking University, Beijing 100871, China} 
\affiliation[c]{CAS Center for Excellence in Particle Physics, Beijing 100049, China} 
\affiliation[d]{Institute of Physical Science and Information Technology, Anhui University, Hefei 230026, China}

\emailAdd{zuoweiliu@nju.edu.cn} 
\emailAdd{mg1622019@smail.nju.edu.cn} 
\emailAdd{dayu@nju.edu.cn}

\abstract{
We investigate the capability of the future electron collider CEPC in 
probing the parameter space of several dark matter models, 
including millicharged dark matter models, $Z'$ portal dark matter models, 
and effective dark matter operators.
In our analysis, the monophoton final state is used as 
the primary channel to detect dark matter models at CEPC. 
To maximize the signal to background significance, 
we study the energy and angular distributions 
of the monophoton channel arising from dark matter models 
and from the standard model 
to design a set of detector cuts. 
For the $Z'$ portal dark matter, we also analyze the 
$Z'$ boson visible decay channel which is found to be complementary 
to the monophoton channel in certain parameter space. 
The CEPC reach in the parameter space of dark matter models 
is also put in comparison with Xenon1T. 
We find that CEPC has the unprecedented sensitivity 
to certain parameter space for the dark matter models considered; 
for example, CEPC can improve the limits on millicharge 
by one order of magnitude than previous collider experiments 
for ${\cal O}(1)-100$ GeV dark matter.

}

\begin{document}
\maketitle
\flushbottom

\section{Introduction}
\label{sec:intro}

Astrophysical observations tell us that the 
baryonic matter only contributes about 5\% of the 
energy density of the current universe; more than 80\% of the 
matter content consists of unknown dark matter (DM)
particles \cite{Akrami:2018vks}. 
To investigate the particle nature of DM is one of the 
pressing issues in new physics studies beyond 
the standard model (SM).  
There is a large variety of experiments in which one can 
probe the DM particle properties, including dark matter 
direct detection experiments, 
dark matter indirect detection experiments, 
and particle colliders. 
In this paper, we study the capability of the 
proposed circular electron positron collider (CEPC) 
in probing various dark matter models. 

Three different running modes for CEPC 
have been proposed \cite{CEPCStudyGroup:2018ghi}, 
including the Higgs factory mode (hereafter the $H$-mode) 
with $\sqrt{s} = 240$ GeV for the 
$e^+e^- \to ZH$ production 
and a total luminosity 
of $\sim$5.6 ab$^{-1}$ for seven years,  
the $Z$ factory mode (hereafter the $Z$-mode) 
with $\sqrt{s}=91.2 \GeV$ 
for the $e^+e^- \to Z$ production 
and a total luminosity 
of $\sim$16 ab$^{-1}$ for two years,  
and the $WW$ threshold scan  (hereafter the $WW$-mode) 
with $\sqrt{s} \sim 158-172\GeV $ 
for the $e^+e^- \to W^+W^-$ production
and a total luminosity 
of $\sim$2.6 ab$^{-1}$ for one year \footnote{We take 
$\sqrt{s}=160$ GeV for the $WW$-mode throughout our analysis.}. 
The unprecedented luminosity and energy of CEPC will 
enable physicists to study the unexplored   
region in both SM and 
new physics beyond SM. 

Besides CEPC, several other future lepton colliders 
have been proposed, including the 
International Linear Collider (ILC) 
\cite{Baer:2013cma}, 
the Future Circular Collider of $e^+e^-$ (FCC-ee) 
\cite{Gomez-Ceballos:2013zzn}, 
and the Compact Linear Collider (CLIC) 
\cite{Abramowicz:2013tzc}. 
These new lepton colliders will certainly 
deepen our understandings about the dark matter 
or hidden sector 
(see e.g.\ \cite{Birkedal:2004xn,Konar:2009ae,Fox:2011fx,
Bartels:2012ex,Chae:2012bq,Dreiner:2012xm,
Yu:2013aca,Yu:2014ula,Neng:2014mga,Karliner:2015tga,
Harigaya:2015yaa,Cao:2016qgc,Xiang:2016jni,Cai:2016sjz,
Cai:2017wdu,Xiang:2017yfs,Wang:2017sxx,He:2017ord,
He:2017zzr,Liu:2017zdh,Gao:2017tgx,Kadota:2018lrt}
for some recent studies).

In this work, we investigate the collider signatures of 
millicharged DM models which have not been studied 
at CEPC. 
We further explore the constraining power of 
CEPC in probing the $Z'$ portal DM models from the 
ordinary dark matter channels, as well as 
from $Z'$ visible decay searches. 
We also study the collider signals arising from 
dark matter models in which dark matter 
interacts with SM via effective field theory operators. 
We study the monophoton signatures at CEPC for 
the DM related processes and the relevant SM processes. 
The CEPC upper bounds on dark matter processes as well as
on the coupling strength are analyzed for the three proposed 
running modes. We find that CEPC has the potential to probe 
the parameter space that is currently unexplored by previous experiments  
in various dark matter models.

The rest of the paper is organized as follows. 
In section \ref{sec:model}, 
we introduce the dark matter models that are   
investigated in this paper. 
In section \ref{sec:signal-bg}, 
we study the monophoton signature  
arising from DM models 
and from the SM. 
A set of detector cuts 
to suppress the SM backgrounds 
and to maximize the signal-to-background ratio 
are proposed. 
We present the results for millicharged models, 
$Z'$ portal models, and DM effective operators 
in section \ref{sec:milliq}, section \ref{sec:zp} and section \ref{sec:eft} respectively. 
A preliminary study on automatic optimizations of detector cuts 
is given in section \ref{sec:opt}. We summarize our findings in section \ref{sec:sum}.


\section{Dark matter models}
\label{sec:model}

We consider the following three types of dark matter models: 
(1) millicharged DM; 
(2) $Z'$ portal DM; 
(3) DM interactions with SM via effective-field-theory (EFT) operators. 
The interaction Lagrangian of the millicharged DM is given by 
\begin{equation}
{\cal L} = e \varepsilon A_\mu \bar \chi \gamma^\mu \chi, 
\label{eq:millicharge}
\end{equation}
where $\chi$ is a millicharged Dirac DM particle, 
$A_\mu$ is the SM photon, 
$e$ is the electromagnetic coupling strength,  
and $\varepsilon$ is the millicharge. 
The interaction Lagrangian for the $Z'$ portal DM model is given by 
\be
{\cal L} = Z'_\mu\bar{\chi}\gamma^\mu(g^\chi_V-g^\chi_A\gamma_5)\chi 
+ Z'_\mu\bar{f}\gamma^\mu(g^f_V-g^f_A\gamma_5)f,
\ee
where $\chi$ is the Dirac DM, 
$f$ is the SM fermion,
$Z'$ is a spin-1 particle that interacts with DM and with SM fermions. 
We consider both vector and axial-vector couplings between 
the $Z'$ boson and fermions. 
There are a variety of EFT operators that one can introduce to mediate  
the interaction between DM and the SM particles. 
Here we consider the following four-fermion effective 
field theory operators \cite{Fox:2011fx} \cite{Chae:2012bq} 
\begin{equation}
\label{eq:eftl}
\begin{aligned}
	{\cal L}=& \frac{1}{\Lambda_V^2} \bar{\chi}\gamma_\mu \chi \bar{\ell}\gamma^\mu \ell,\\
	{\cal L}=& \frac{1}{\Lambda_s^2} \bar{\chi}\chi \bar{\ell} \ell,\\
	{\cal L}=& \frac{1}{\Lambda_A^2} \bar{\chi}\gamma_\mu\gamma_5\chi\bar{\ell}\gamma^\mu\gamma_5 \ell,\\
	{\cal L}=& \frac{1}{\Lambda_t^2} \bar{\chi} \ell \bar{ \ell } \chi
\end{aligned}
\end{equation}    
where $\chi$ is the Dirac DM, 
$\ell$ denotes the SM charged lepton, 
and the various $\Lambda$ 
parameters are the characteristic scales for new physics. 
Here $\Lambda_V$ ($\Lambda_A$) is the 
new physics scale for vector (axial-vector) interaction; 
$\Lambda_s$ ($\Lambda_t$) is the scalar 
interaction which can be obtained 
with an $s-$channel ($t-$channel) mediator 
integrated out. 
We note in passing that the effective operator approach 
starts to 
break down when the momentum transfer becomes 
comparable to the suppression scale $\Lambda$.


\section{Signals and backgrounds}
\label{sec:signal-bg}


\subsection{DM signals}
\label{subsec:signal}

\begin{figure}[htbp]
\begin{centering}
\includegraphics[scale=1.1]{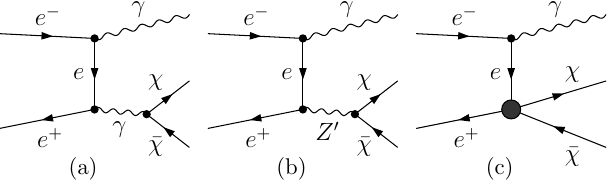}
\caption{Feynman diagrams for the process \eexxa
in the millicharged DM models (a), 
in the $Z^\prime$-portal DM models (b), 
and in the DM effective operator formalism (c).
Diagrams with photon radiated by the positron are 
included in the physics analysis but not shown here.
The diagrams with photon emitted by the millicharged 
DM are neglected due to the small $\varepsilon$ value.}
\label{fig:monophoton-fd}
\end{centering}	
\end{figure}

Typically, dark matter escapes the particle detectors 
without leaving any directly detectable 
signal.\footnote{Millicharged particles can be detected with very sensitive 
detectors; however, for the typical detectors in the high energy 
colliders, the millicharged particles remain unseen.}
Thus, in order to detect DM particles in colliders, 
at least one final state visible particle is required to be produced 
in association with the final state DM particles. 
Here we use the monophoton signature to 
probe the DM models. 
Fig.\ (\ref{fig:monophoton-fd}) shows the Feynman diagrams for 
the production process of DM particles in association with a 
single photon in the final state for the millicharged DM models, 
the $Z^\prime$ portal DM models, and DM effective operators.

For the milicharged DM models, 
the differential cross section for the \eexxa
process is given by \cite{Liu:2018jdi} 
\be
{d \sigma \over d E_\gamma d z_\gamma} = 
{8 \alpha^3 \varepsilon^2 (1 + 2 y) \beta_\chi
\over 3 s E_\gamma} 
\Bigg[{1  + x (1+z_\gamma^2) 
\over 1 - z_\gamma^2 }
\Bigg], 
\label{eq:mq}
\ee
where 
$s$ is the center-of-mass energy square, 
$m_\chi$ is the DM mass, 
$E_\gamma$ is the final state photon energy, 
$s_\gamma = s - 2\sqrt{s} E_\gamma$,  
$y \equiv m_\chi^2 / s_\gamma$,  
$x \equiv E_\gamma^2 / s_\gamma$, 
$\beta_\chi = (1-4 y)^{1/2}$, 
and $z_\gamma \equiv \cos\theta_\gamma$ with   
$\theta_\gamma$ being the polar angle of the final state photon. 


For the $Z'$ portal DM models where both vector 
and axial-vector 
couplings to fermions are present, 
the production cross section for the process 
$e^+ e^- \to \gamma Z^\prime \to \gamma \chi \bar \chi$ 
is given by 
\be
{d \sigma \over d E_\gamma d z_\gamma} = 
{\alpha 
\Big[ (g^f_{V})^2  + (g^f_{A})^2 \Big]
\Big[ (g^\chi_{V})^2 (1+2y)+(g^\chi_{A})^2 (1-4y) \Big]
s_\gamma^2 \beta_\chi 
\over 6 \pi^2 s E_\gamma
\left[ 
(s_\gamma - M_{Z^\prime}^2)^2 + M_{Z^\prime}^2 \Gamma_{Z^\prime}^2
\right] 
}  
\Bigg[{1  + x (1+z_\gamma^2) 
\over 1 - z_\gamma^2 }
\Bigg], 
\label{eq:zpVAV}
\ee
where $M_{Z^\prime}$ is the $Z^\prime$ mass, 
$\Gamma_{Z'}$ is the total $Z'$ decay width, 
which is given by 
\be
\Gamma_{Z'} = \Gamma(Z' \to \chi \bar \chi) + \sum_f \Gamma (Z' \to f \bar f). 
\ee
The DM decay width is given by 
\begin{equation}
\Gamma(Z' \to \chi \bar \chi)= \frac{M_{Z'}}{12\pi}
\sqrt{1-4 {m_\chi^2 \over M_{Z'}^2}}
\Bigg[
(g_V^\chi)^2 \Big(1+2 {m_\chi^2 \over M_{Z'}^2} \Big)
+
(g_A^\chi)^2 \Big(1-4 {m_\chi^2 \over M_{Z'}^2} \Big)
\Bigg]. 
\end{equation}
The decay widths into SM fermions can be 
obtained by substituting the couplings 
and mass by the SM values. 
The production cross sections for the 
DM effective operators 
can be found in Ref.\ \cite{Chae:2012bq}.

The maximum energy that the final state photon can have 
is given by 
\be
E_\gamma <  {s - 4 m_\chi^2 \over 2 \sqrt{s} } 
\equiv E_\chi^{m}. 
\ee
For the $Z'$ portal DM models, 
the monophoton energy spectrum exhibits a resonance 
centered at the photon energy 
\be
E_\gamma = {s - M_{Z'}^2 \over 2 \sqrt{s} }, 
\ee 
with a full-width-at-half-maximum (FWHM) as 
$(M_{Z'}/\sqrt{s})\Gamma_{Z'}$, 
due to the Breit-Wigner distribution of the $Z'$ boson. 
We will refer to such a peak as the ``$Z'$ resonance'' hereafter. 
For the case where $2 m_\chi > M_{Z'}$, 
the resonance in the photon energy spectrum 
exceeds the maximum energy 
of the final state photon and thus cannot be observed. 



\begin{figure}[htbp]
\begin{centering}
\includegraphics[width=0.45\columnwidth]{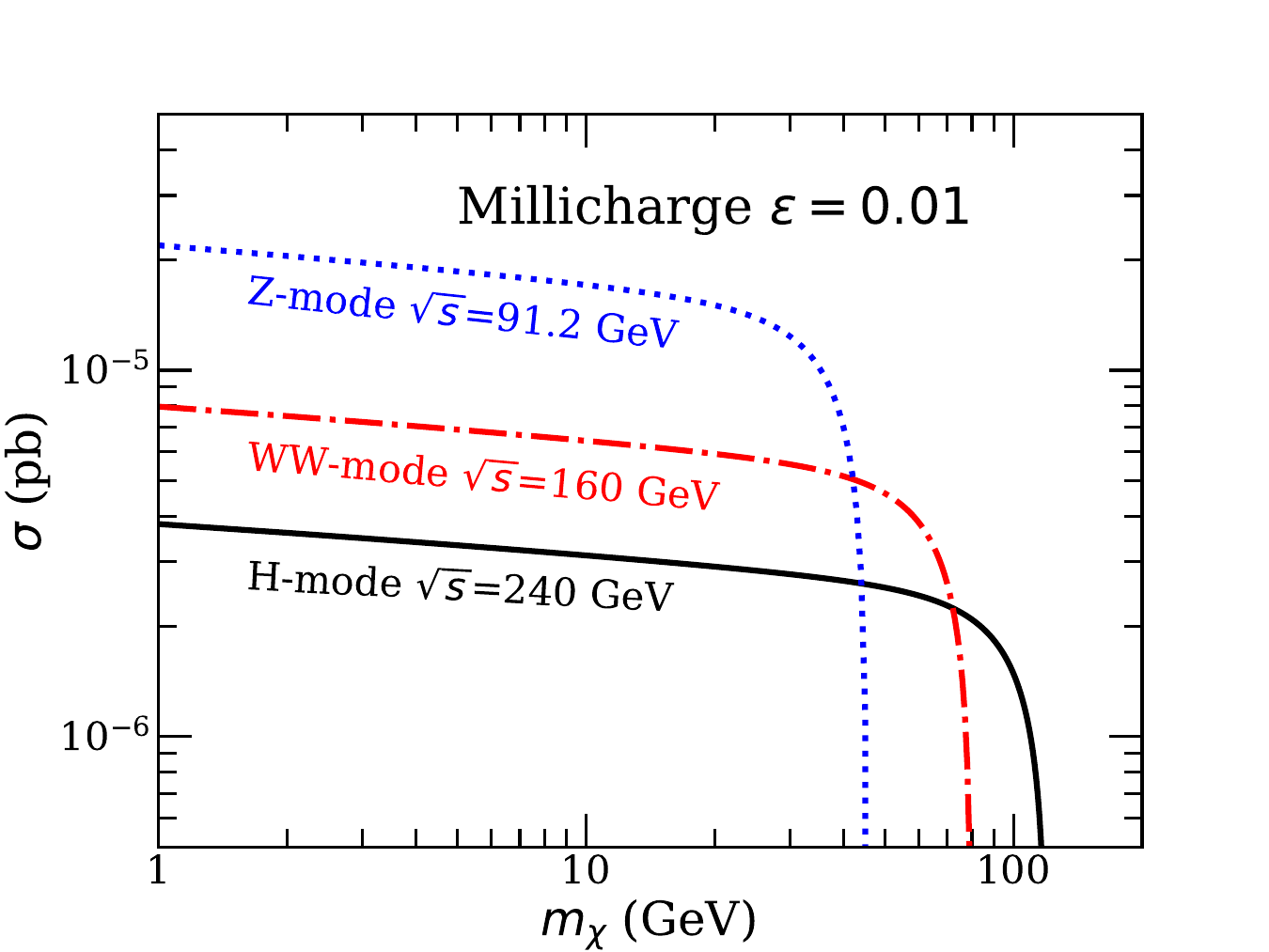}
\includegraphics[width=0.45\columnwidth]{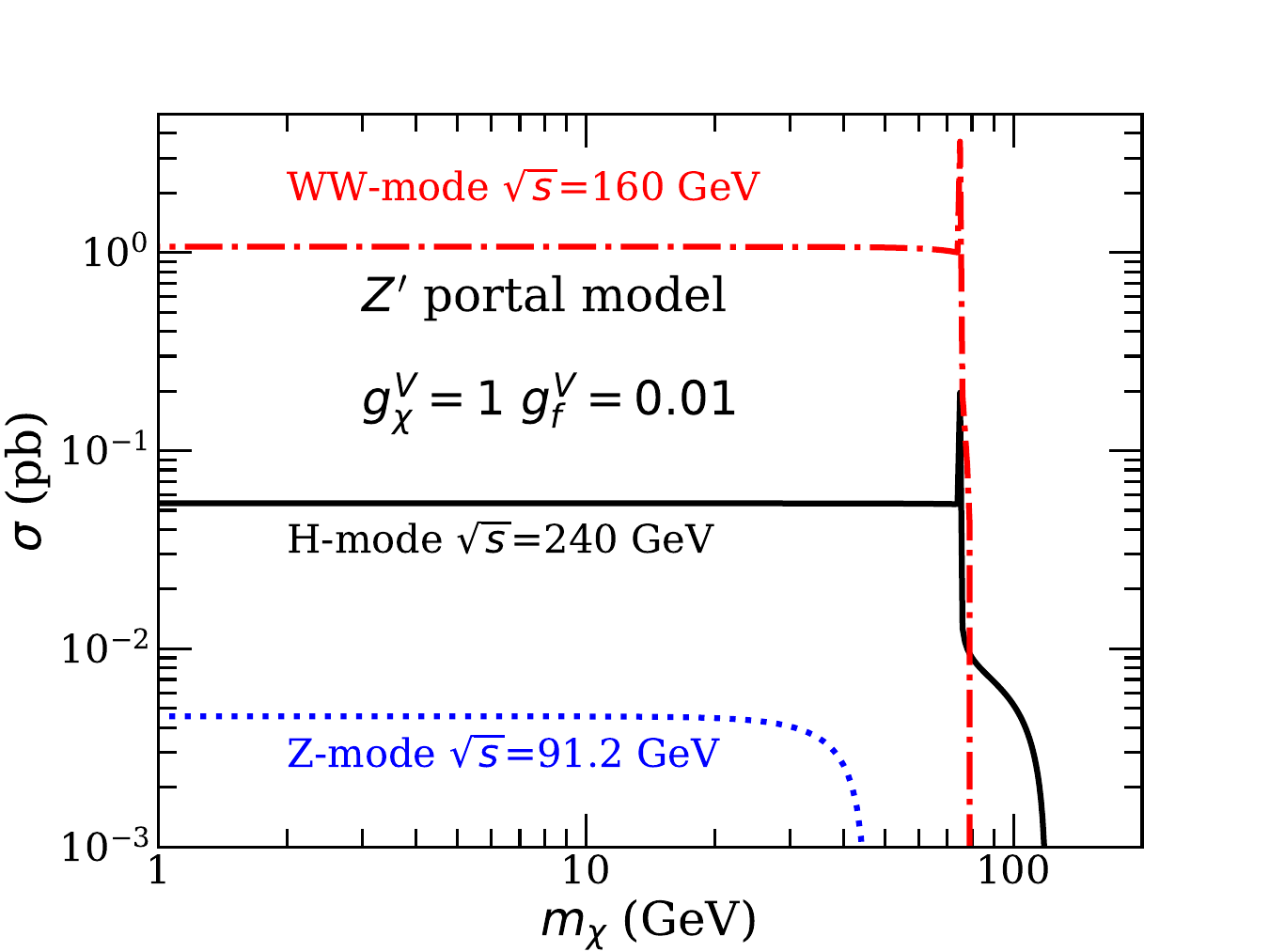}
\includegraphics[width=0.45\columnwidth]{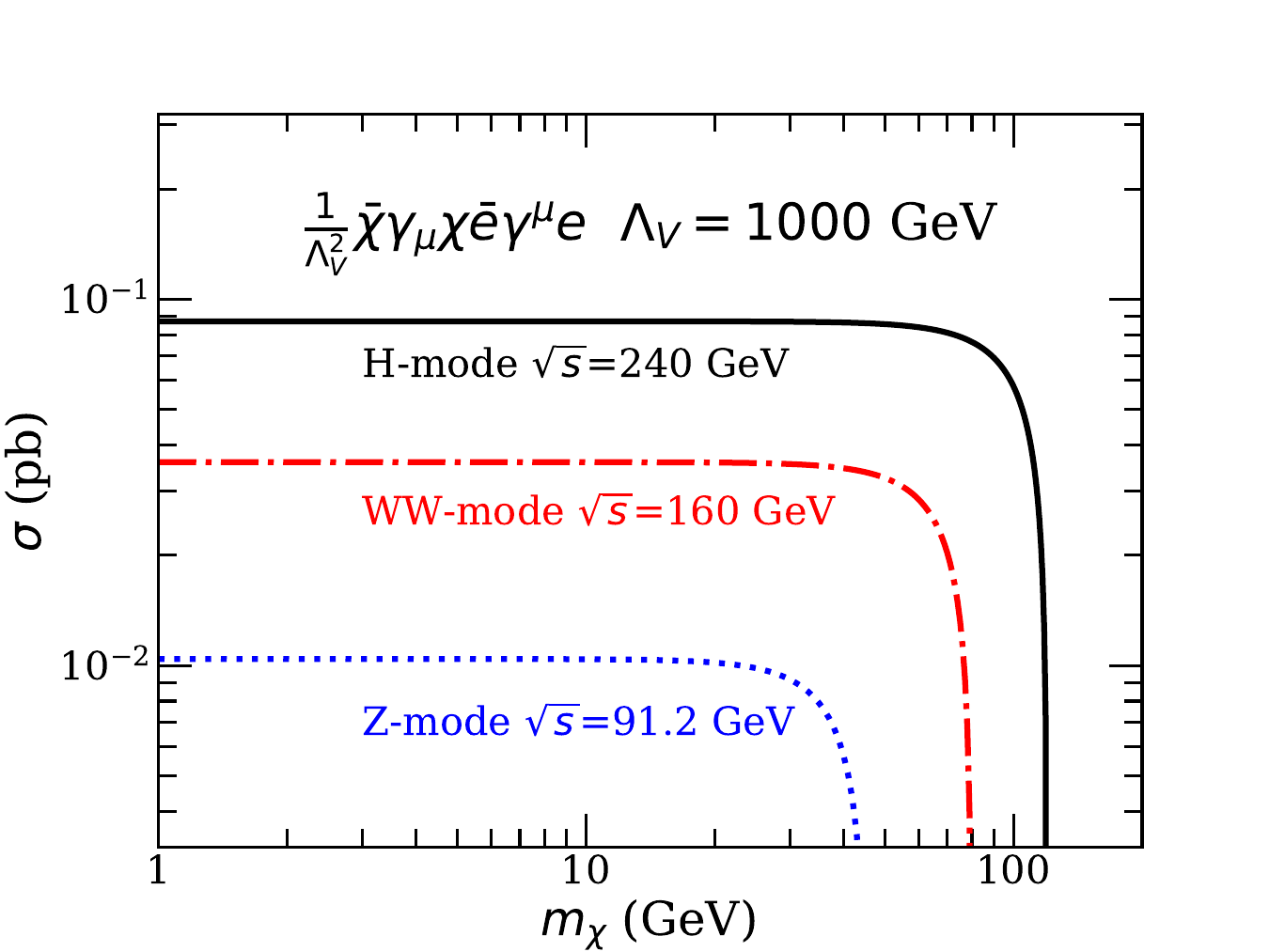}
\includegraphics[width=0.45\columnwidth]{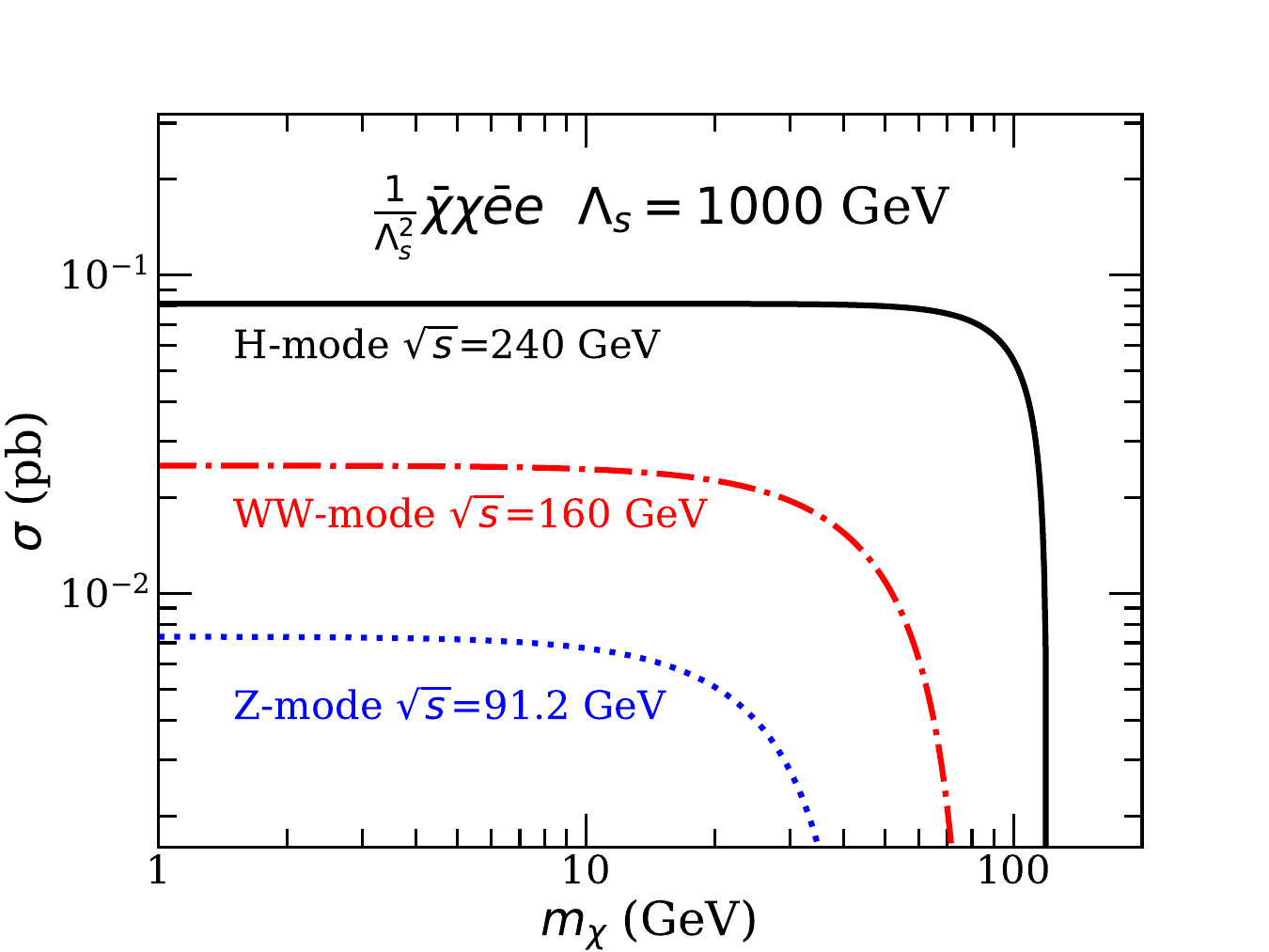}
\caption{
Total monophoton cross section 
$\sigma(e^+e^-\to \chi \bar \chi\gamma)$ 
at CEPC in millicharged DM models (upper left panel), 
in $Z'$ portal DM models (upper right panel), 
and with DM effective operators (lower two panels). 
For the millicharged DM models, we use $\varepsilon = 0.01$ here. 
For the $Z'$ portal DM models, we use 
$M_{Z'} = 150\GeV$, $g_V^f = 0.01$, and $g_V^\chi = 1$ 
in the vector coupling only case. 
For the DM effective operators, we use $\Lambda = 1000\ \GeV$.  
The three different running modes are considered here: 
the $H$-mode, the $Z$-mode, and the $WW$-mode. 
The monophoton cross sections are computed using the 
detector cuts:  $E_\gamma > 0.1 \GeV$ and $|z_\gamma|<0.99$.
}	
\label{fig:mqxsections}
\end{centering}
\end{figure}


In Fig.\ (\ref{fig:mqxsections}), 
we compute the total monophoton 
cross section at CEPC for the three dark matter models, 
by integrating the differential cross section over the region: 
$E_\gamma>0.1$ GeV and $|z_\gamma|<0.99$. 
The center-of-mass energy square is 
$\sqrt{s} = 240$ (91.2) GeV for the 
$H$-mode ($Z$-mode); 
for the $WW$-mode, we use $\sqrt{s}=160 \GeV$ 
as the benchmark point.

In the millicharged DM model, the monophoton cross 
section increases when $\sqrt{s}$ decreases at CEPC, 
for DM mass lighter than 40 GeV, 
as shown in the upper-left panel figure of Fig.\ (\ref{fig:mqxsections}). 
Thus the $Z$-mode has the better sensitivity than the other two modes in 
probing light millicharged DM. 
For the four-fermion effective operators, 
the monophoton cross section increases 
when $\sqrt{s}$ increases at CEPC, 
as shown in the lower two panel figures of Fig.\ (\ref{fig:mqxsections}). 
Thus the $H$-mode has the better sensitivity than the other two 
low energy modes in 
probing the DM effective operators with TeV suppression scale.

For the $Z'$ portal DM model, we consider the case in which 
the mass of the 
$Z'$ boson is 150 GeV; 
the monophoton cross section in the $WW$-mode is 
larger than the other two modes, 
for the case in which DM is lighter than $\sim 70$ GeV, 
as shown in the upper-right panel figure of Fig.\ (\ref{fig:mqxsections}), 
where we consider the vector coupling only case. 
It is interesting to note that the monophoton cross section 
exhibits a resonance feature when the mass of the 
DM is in the vicinity of  $M_{Z'}/2$. 
Thus we expect a better sensitivity in probing the 
$Z'$ portal DM models in the $WW$-mode, and 
enhanced constraints in the parameter space 
where the $Z'$ boson is nearly twice of the DM mass. 
We note that the different relations between the total 
production cross section and $\sqrt{s}$ in the three 
types of DM models are primarily 
due to the mass scale of the mediator. 



\subsection{SM backgrounds}
\label{sec:sm} 

We discuss the major SM backgrounds (BG) at CEPC to the 
monophoton signature, which include the irreducible background 
and reducible background.

\begin{figure}[htbp]
\centerline{\includegraphics[width=11cm]{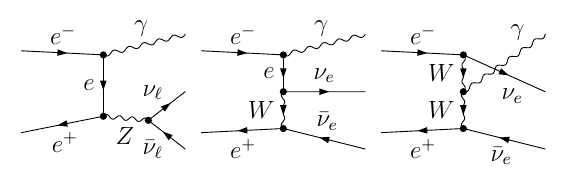}}
\caption{Leading order $e^+ e^-\to \nu \bar\nu \gamma$ processes at CEPC. 
The $W$-boson can mediate the 
$e^+e^-\rightarrow \nu_e\bar{\nu_e}\gamma$ processes; 
the $Z$-boson diagrams contribute to all neutrino flavors. 
}
\label{fig:feynman-irbg}
\end{figure}

The irreducible SM background to the 
monophoton signature at CEPC are the 
$e^+e^-\rightarrow\nu_\ell\bar{\nu_\ell}\gamma$ processes,
where $\nu_\ell = \nu_e,\,\nu_\mu,\,\nu_\tau$ 
are the three SM neutrinos.
The corresponding Feynman diagrams are 
displayed in Fig.\ (\ref{fig:feynman-irbg}). 
For muon and tau neutrinos, only $Z$-boson diagrams 
contribute; for the electron neutrino, both $Z$-boson 
diagrams and $W$-boson diagrams contribute.

\begin{figure}[htbp]
\begin{centering}
\includegraphics[width=0.45\columnwidth]{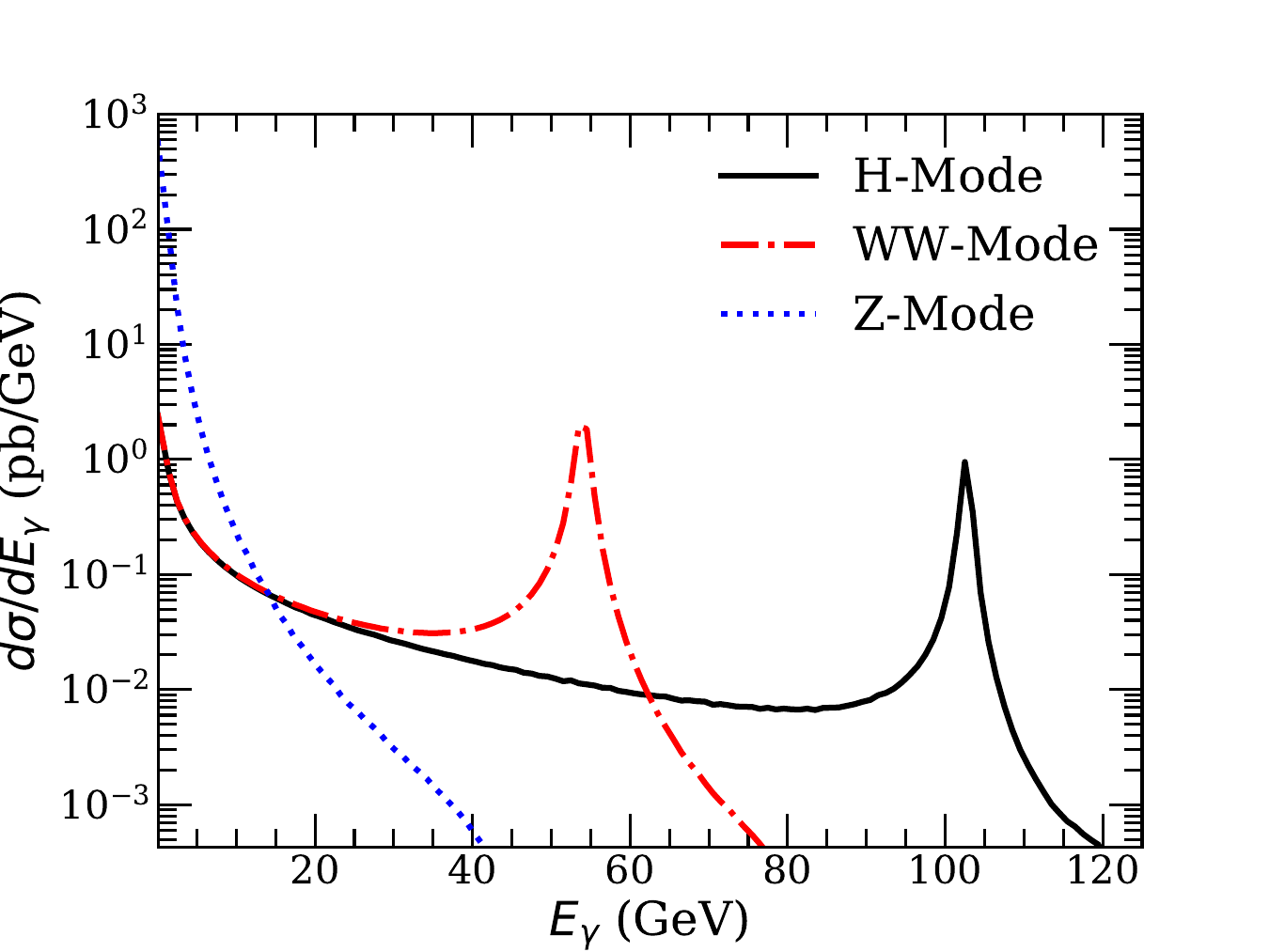}
\caption{Differential cross section for the irreducible standard model background 
process $e^+ e^- \to \nu \bar \nu \gamma$. 
We compute the SM cross sections using {\sc FeynAtrs} \cite{Hahn:2000kx} 
and {\sc FormCalc} \cite{Hahn:1998yk} packages
and consider the three CEPC modes with the 
detector cuts $E_\gamma > 0.1\GeV$ and $|\cos\theta_\gamma| < 0.99$. 
}
\label{fig:xsec-irbg}
\end{centering}
\end{figure}

Due to the SM $Z$ boson, 
the irreducible BG exhibits a resonance 
in the monophoton energy spectrum which is 
centered at the photon energy 
\be
E^Z_\gamma = {s - M_{Z}^2 \over 2 \sqrt{s} }
\ee 
with a FWHM as 
$\Gamma^Z_\gamma = M_{Z} \Gamma_{Z} /\sqrt{s}$. 
We will refer to this resonance in the monophoton 
energy spectrum as the ``$Z$ resonance'' hereafter. 
For the three CEPC running modes, the resonance 
in the monophoton energy spectrum are located at 
$E^Z_\gamma \simeq 103$ GeV 
with $\Gamma^Z_\gamma \simeq 0.95$ GeV for the $H$-mode, 
$E^Z_\gamma \simeq 54$ GeV 
with $\Gamma^Z_\gamma \simeq 1.4$ GeV for the $WW$-mode 
(for $\sqrt{s}=160$ GeV), 
and 
$E^Z_\gamma \simeq 0$ GeV 
with $\Gamma^Z_\gamma \simeq 2.5$ GeV for the $Z$-mode. 
The differential cross section of the monophoton channel due to 
the SM irreducible background is shown in Fig.\ (\ref{fig:xsec-irbg}) 
where the detector effects have not been taken into account. 
The $Z$ resonance features can be easily visualized in the 
monophoton energy spectrum as shown in Fig.\ (\ref{fig:xsec-irbg}). 
Thus, we will veto the events within $5\Gamma^Z_\gamma$ 
at the $Z$ resonance in the monophoton energy spectrum 
to suppress the irreducible background contribution.

Next we investigate the reducible background for the monophoton signature 
at the CEPC. 
The reducible backgrounds arise due to the limited detection
capability of the detectors. 
Following CEPC CDR \cite{CEPCStudyGroup:2018ghi}, 
we adopt the following parameters for the EMC coverage:  
$|\cos\theta| < 0.99$ and $E>0.1$ GeV. 
For the photon energy resolution, we use the 
resolution in the dual-readout calorimeter \cite{CEPCStudyGroup:2018ghi} 
\be
{\sigma(E) \over E} = 
\frac{10.1 \%}{\sqrt{E/{\rm GeV}}}\bigoplus0.4\%. 
\label{eq:resolution}
\ee
We use the above detector parameters in our simulations. 
The EMC positional resolution has been estimated as $\sim 0.1$ mm \cite{Wang::2017}, 
which gives rise to a relative spatial resolution as $\sim 10^{-5}$, since the 
radius of the detector cylinder is about 2 meters. 
Thus we do not take the spatial resolution into consideration in 
our simulations due to its smallness.  

\begin{figure}[htbp]
\begin{centering}
\includegraphics[width=0.5\columnwidth]{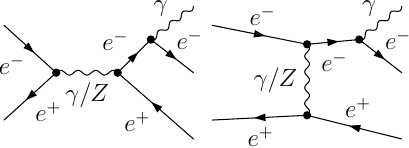}
\caption{Tree level diagrams for the 
$e^+e^-\rightarrow e^+e^-\gamma$ processes in SM.}
\label{fig:feynman-rbg}
\end{centering}
\end{figure}

Thus the major reducible
SM backgrounds come from the $e^+e^-\to \gamma +\slashed{X}$ processes, 
where only one particle in the final state is visible to the particle detectors which 
is the final state photon, 
and $\slashed{X}$ denotes the other particle (or particles) in the final state 
that are undetected due to the limitations of the detectors. 
In SM, the dominate reducible backgrounds include the processes in which 
$X=\bar f f$ where $f$ is an SM fermion, $X=\gamma$, and $X=\gamma \gamma$. 
The contribution to the monophoton signal from the 
$e^+e^-\rightarrow\gamma\gamma$ 
process is negligible because the CEPC detectors 
are arranged in a symmetric manner. 
However, the reducible background from the processes 
$e^+e^-\rightarrow f\bar{f}\gamma$ 
and 
$e^+e^-\rightarrow \gamma \gamma \gamma$ 
can be quite large 
when the $f\bar{f}$ and $\gamma \gamma$ are 
emitted with $|\cos\theta| > 0.99$. 
For example, 
due to the collinear singularity, 
the $e^+e^-\to \gamma {e}^+ {e}^- $ 
process has large cross section when both final state electron 
and positron go along the beam directions; the corresponding 
Feynman diagrams are exhibited in Fig.\ (\ref{fig:feynman-rbg}). 


\begin{figure}[htbp]
\begin{centering}
\includegraphics[width=0.32\columnwidth]{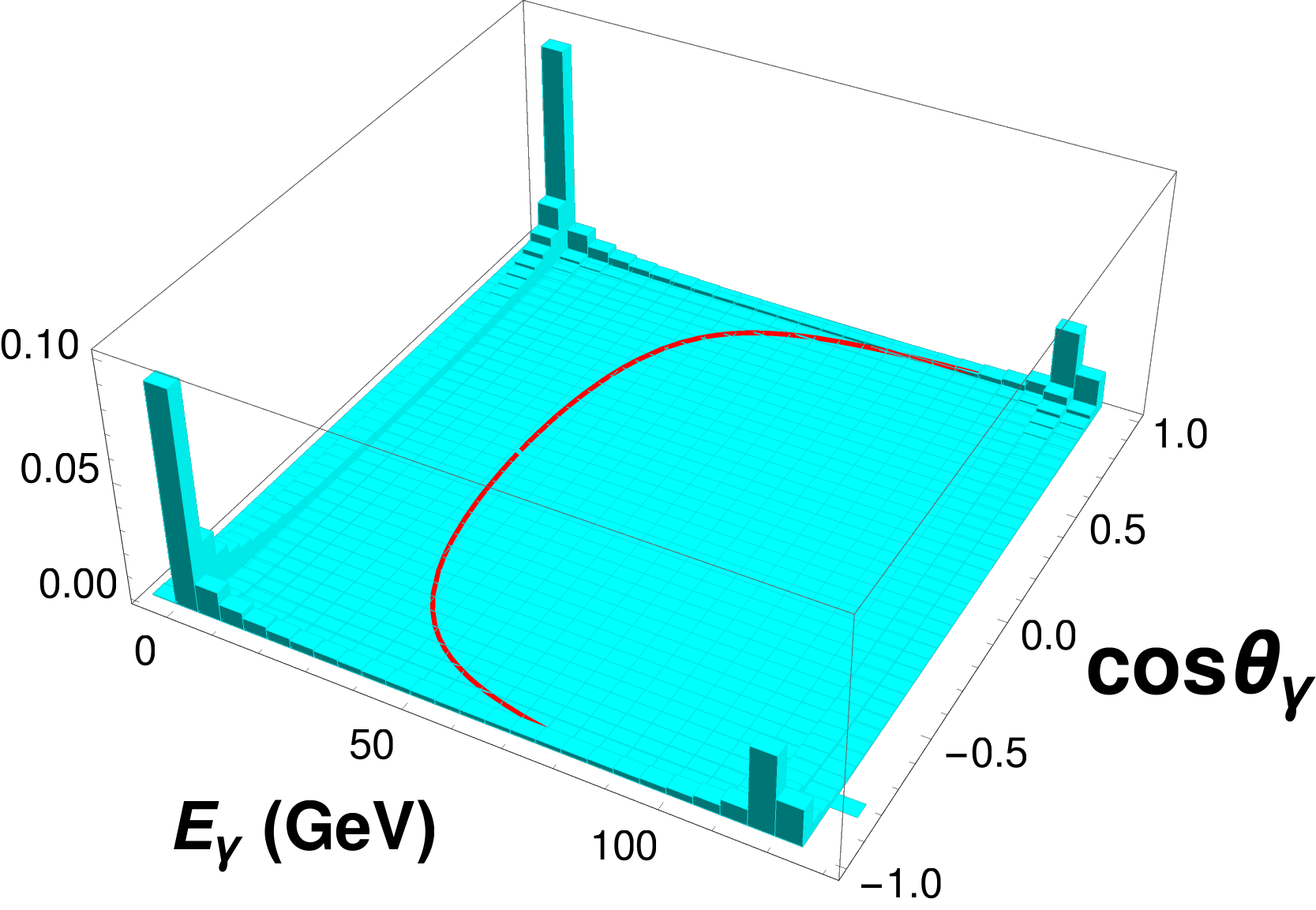}
\includegraphics[width=0.32\columnwidth]{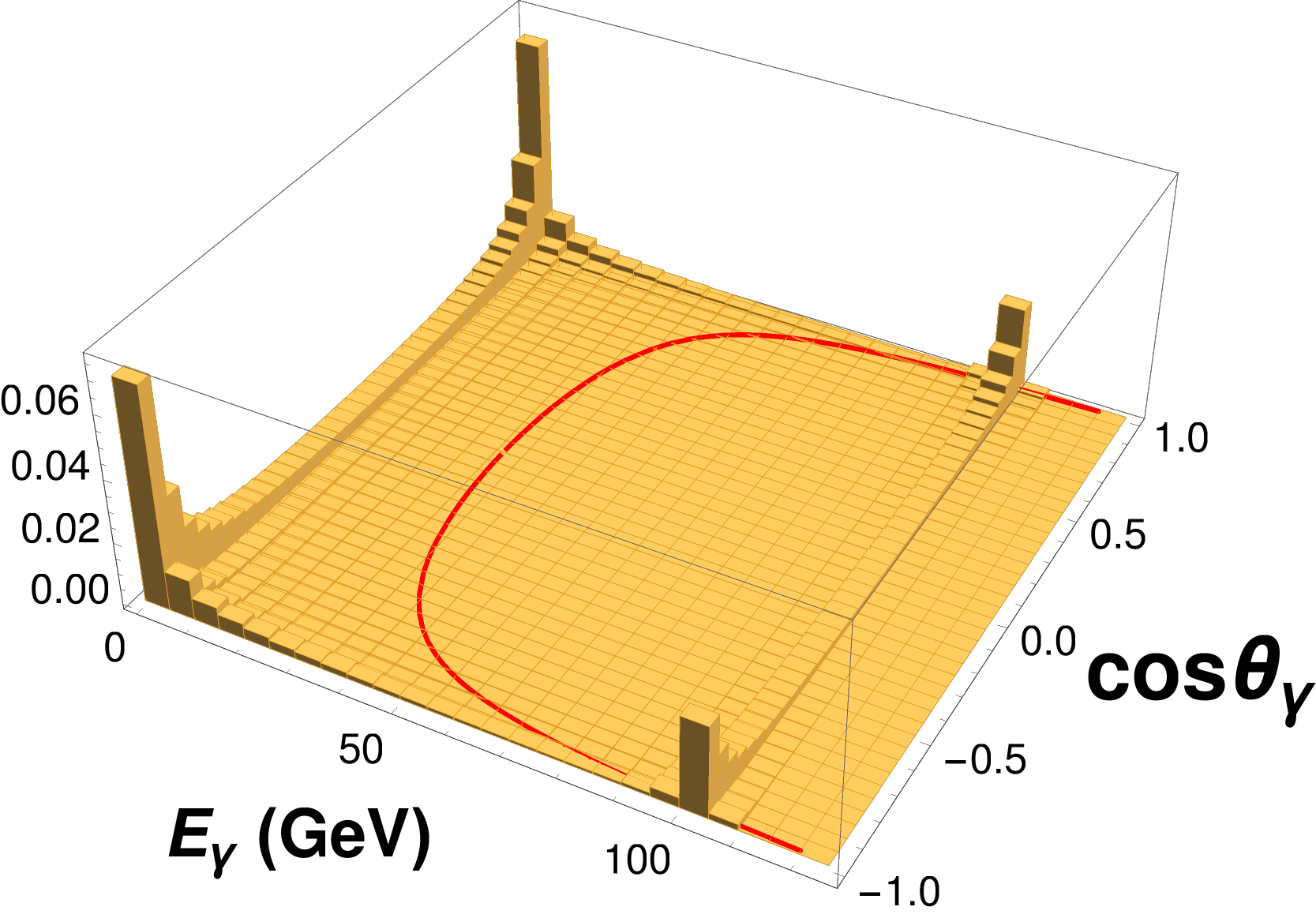}
\includegraphics[width=0.32\columnwidth]{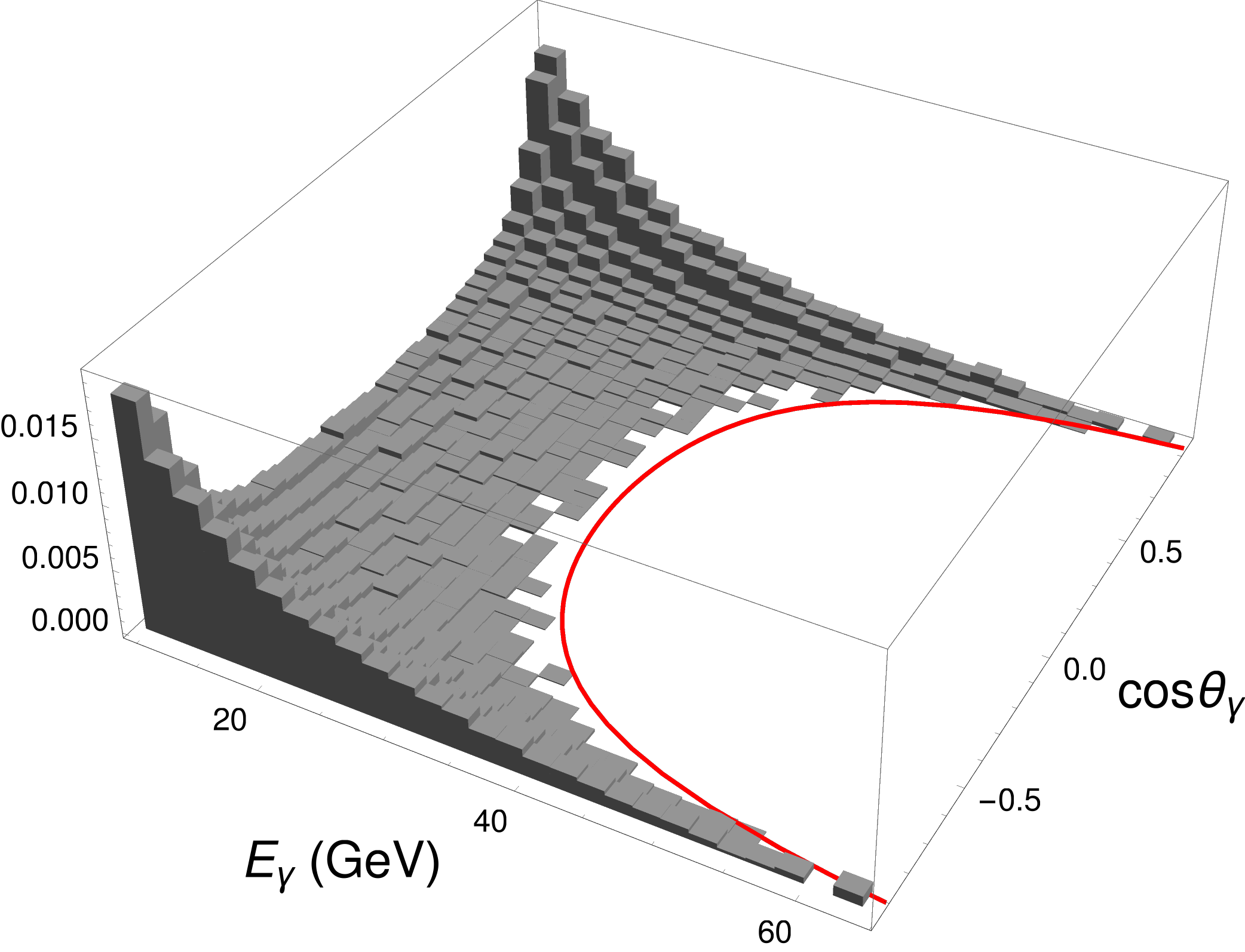}
\caption{Photon $E_\gamma-\cos\theta_\gamma$ distributions 
in millicharged models \eexxa (left) with $m_\chi = 1\GeV$ and $\varepsilon =0.01$,
in SM irreducible BG \eevva (center), 
and in SM reducible BG \eeeea (right). 
$\sqrt{s}=240$ GeV is used here. 
The red curves on each plot indicate the kinematic relation 
$E_\gamma = E_B^{m} = \sqrt{s} (1+\sin\theta_\gamma/\sin\theta_b)^{-1}$ 
where $\cos\theta_b=0.99$. 
The $E_\gamma-\cos\theta_\gamma$ region 
on the right panel plot is 
$E_\gamma > 10\GeV$ \& $|\cos\theta_\gamma| < 0.9$,  
which is different from the first two plots.}
\label{fig:3d}
\end{centering}
\end{figure}


To remove the monophoton events in the reducible 
background, we first compute the energy range 
of the final state photon for certain emitting angle $\theta_\gamma$. 
Below we take the $e^+e^-\to \gamma {e}^+ {e}^- $ as an example. 
We define the polar angle $\theta_{b}$ such that 
$|\cos\theta_{b}|=0.99$ corresponds to the boundary 
of the EMC. 
For certain polar angle $\theta_\gamma$, the 
maximum energy of the photon $E_\gamma^{m}$ 
in the reducible background occurs when the 
electron and positron emit along different beam directions  
with $\theta_{e^\pm} = \theta_{b}$ and 
have transverse momenta $p_T^{e^\pm}$ opposite 
to the photon transverse momentum $p_T^\gamma$. 
By using momentum conservation in the transverse direction 
and energy conservation, we obtain 
$E_\gamma^{m} \sin\theta_\gamma = (\sqrt{s}-E_\gamma^{m}) \sin\theta_{b}$, 
which leads to 
the maximum photon energy as a function of its polar angle as 
\be
E_\gamma^{m} = \sqrt{s} \left[1+ {\sin\theta_\gamma \over \sin\theta_b} \right]^{-1} 
\equiv E_B^{m}(\theta_\gamma).
\label{eq:Em}
\ee
When $\sin\theta_\gamma=1$, 
$E_B^{m}(\theta_\gamma)$ achieves its minimum value 
$(E_B^{m})_{\rm min} \simeq 0.12 \sqrt{s}$. 
For the three CEPC running modes, the minimum value of 
$E_\gamma^{m}(\theta_\gamma)$ 
(at $\theta_\gamma=\pi/2$)  
in the reducible BG is $\sim$29 (19, 11) GeV 
for the $H$ ($WW$, $Z$) mode respectively.  
We thus adopt the detector cut $E_\gamma > E_B^{m}(\theta_\gamma)$ 
on the final state monophoton 
in our analysis to suppress the  
events arising from the SM reducible backgrounds, 
such as the $e^+e^-\rightarrow f\bar{f}\gamma$ 
and $e^+e^-\rightarrow \gamma \gamma \gamma$ 
processes. 
The right panel figure of Fig.\ (\ref{fig:3d}) 
shows that the above cut is efficient 
in removing the reducible backgrounds. 

Thus, to suppress the contributions from SM, 
we apply the following detector cuts to the monophoton 
events in SM and in DM models:  
\begin{itemize}
\item[(1)]  $E_\gamma > 0.1 \GeV$, 
\item[(2)]  $|\cos\theta_\gamma| < |\cos\theta_b| = 0.99$, 
\item[(3)] $E_\gamma < E_\chi^{m} = (s - 4 m_\chi^2)/(2 \sqrt{s})$, 
\item[(4)] veto $E_\gamma \in (E_\gamma^Z \pm 5 \Gamma_\gamma^Z)$, 
\item[(5)] $E_\gamma(\theta_\gamma) > E_{B}^m (\theta_\gamma) 
= \sqrt{s} (1+\sin\theta_\gamma/\sin\theta_b)^{-1}$. 
\end{itemize}
We will collectively refer to the five detector cuts in the list as the ``basic detector cuts'' hereafter. 
Unlike the other detector cuts, the last detector cut in the list is a 2D cut which is applied to 
the two-dimension space spanned by $E_\gamma$ and $\theta_\gamma$. 
Both $E_\gamma^Z$ and $\Gamma_\gamma^Z$ in the 4th detector cut 
are functions of $\sqrt{s}$ in the CEPC running modes.


\section{Millicharged DM models}
\label{sec:milliq}


In this section, we study the monophoton signatures arising from 
the millicharged DM models. 
We use {\sc FeynAtrs} \cite{Hahn:2000kx} and {\sc FormCalc} \cite{Hahn:1998yk} packages
to compute the differential cross sections 
in the millicharged models and in the standard model, 
which are then further smeared using our own code 
to take into account the photon energy resolution.


\begin{figure}[htbp]
\begin{centering}
\includegraphics[width=0.45\columnwidth]{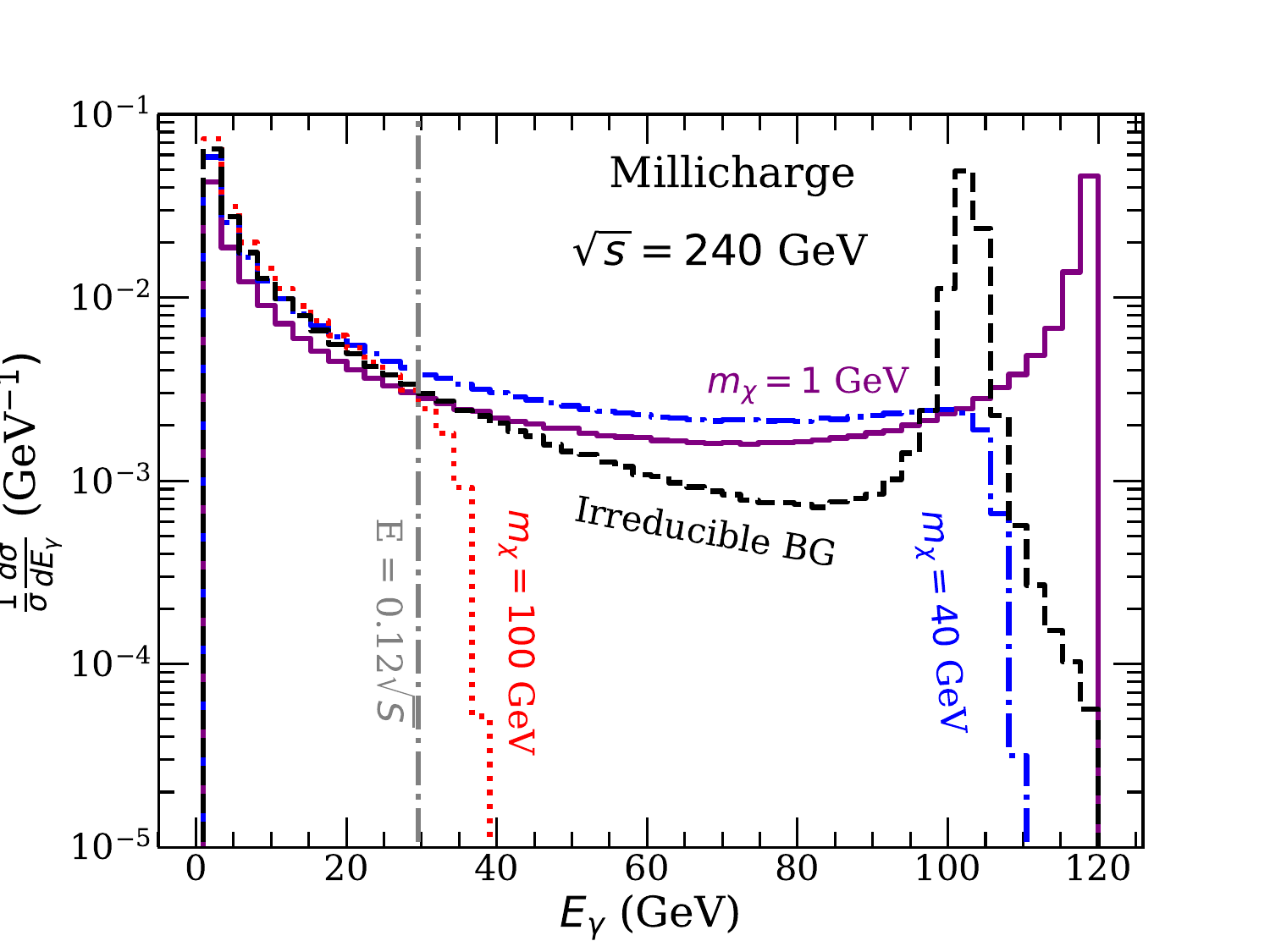}
\includegraphics[width=0.45\columnwidth]{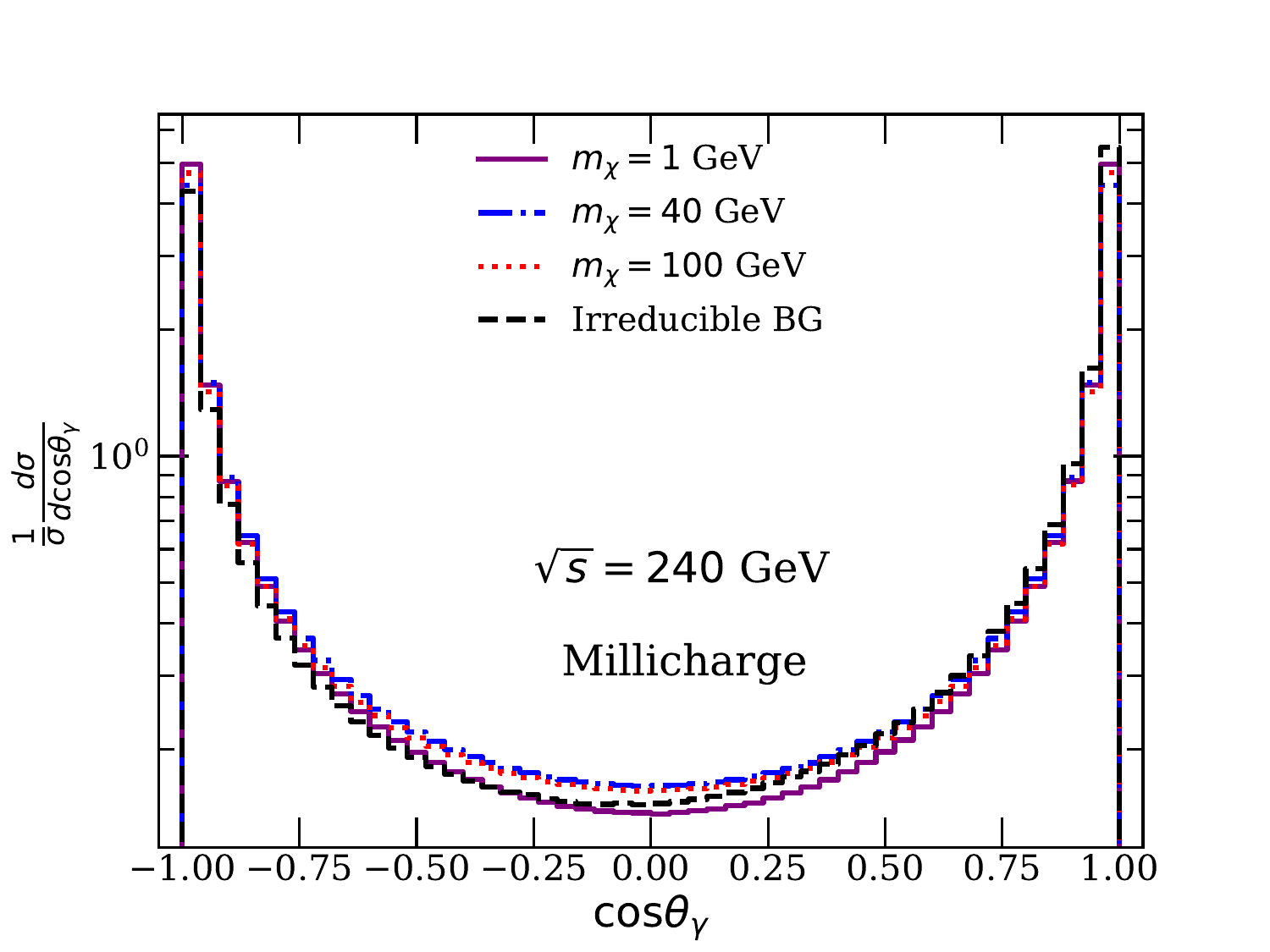}
\caption{Normalized $E_\gamma$ (left) and 
$\cos\theta_\gamma$ (right) 
distributions for the \eexxa processes in millicharged DM models, 
and for the \eevva processes in SM. 
Events in 50 bins with equal width on both plots are computed in the CEPC $H$-mode ($\sqrt{s}=240\GeV$)  
with detector cuts: $E_\gamma > 0.1$ GeV and $|\cos\theta_\gamma| < 0.99$.
For the millicharged models, we consider three different masses: 
$m_\chi=1$ GeV, 40 GeV, and 100 GeV for $\varepsilon = 0.01$. 
The photon energy bin width on the left panel figure, $\sim$ 2.4 GeV, 
is larger than the photon energy resolution which is 
$\delta E_\gamma \simeq 1$ (0.3, 0.1) GeV for $E_\gamma=100$ (10, 1) GeV, 
according to Eq.\ (\ref{eq:resolution}). 
The vertical line $E=0.12 \sqrt{s}$ indicates the boundary of the detector 
cut designed to remove the reducible backgrounds. 
}
\label{fig:mqhist} 
\end{centering}
\end{figure}

\begin{figure}[htbp]
\begin{centering}
\includegraphics[width=0.45\columnwidth]{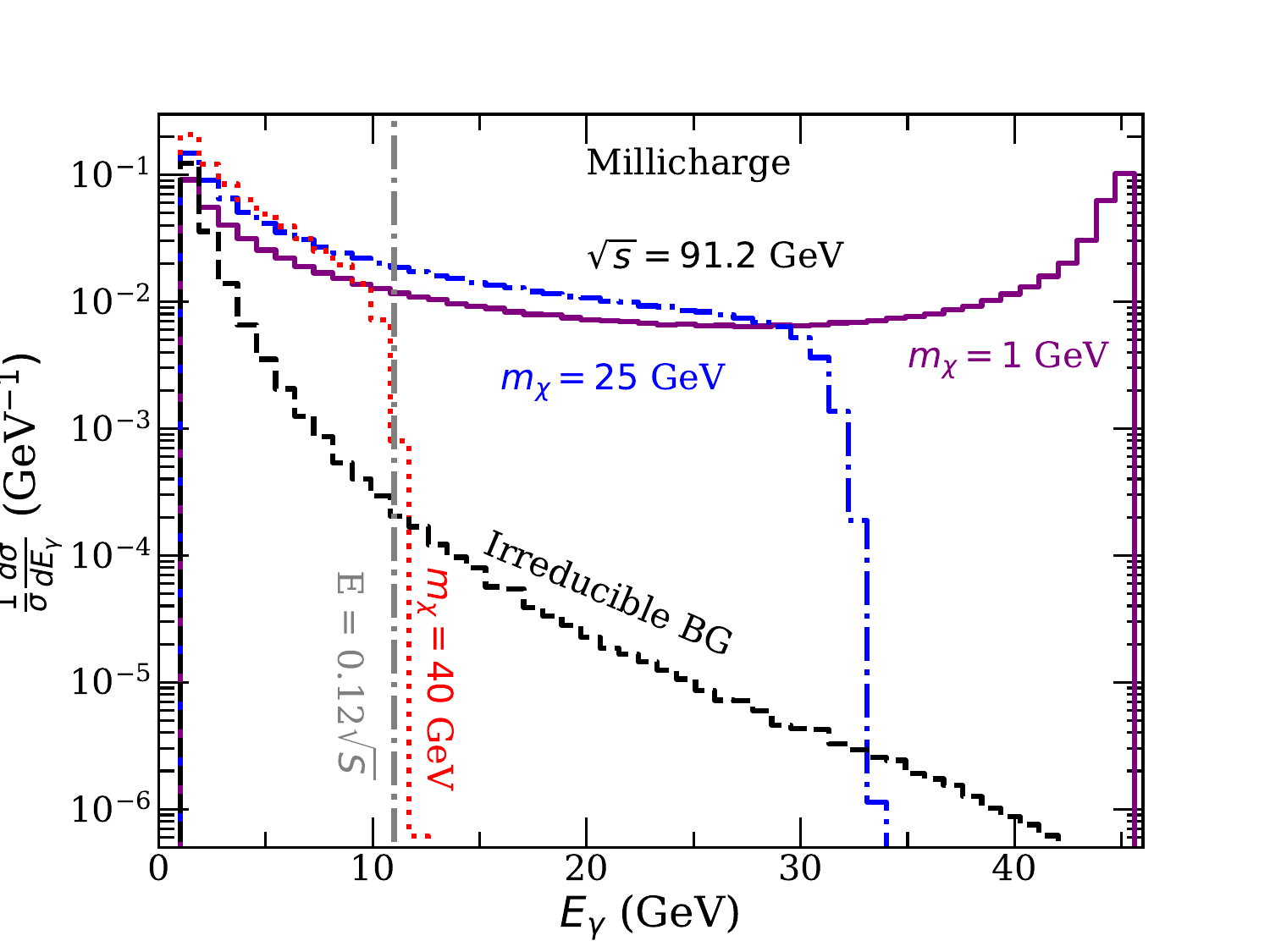}
\includegraphics[width=0.45\columnwidth]{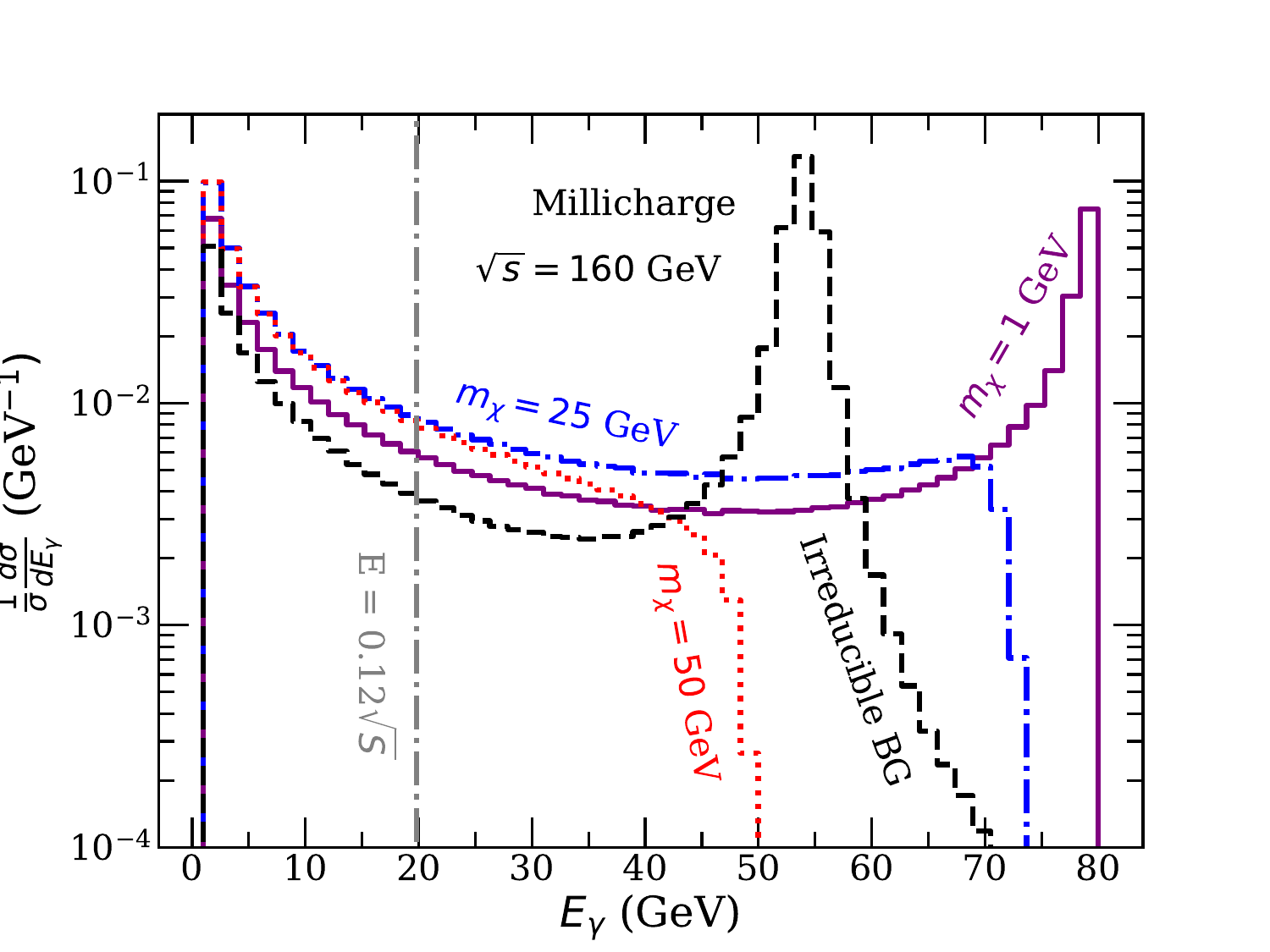}
\caption{Normalized $E_\gamma$ distribution 
in the $Z$-mode (left) 
and in the $WW$-mode (right), 
for both the \eexxa processes in millicharged DM models 
and the \eevva processes in SM. 
Events in 50 bins on both plots are computed 
with detector cuts: $E_\gamma > 0.1$ and $|\cos\theta_\gamma| < 0.99$.
For the millicharged models, we consider $\varepsilon = 0.01$ 
for three different masses in each case; 
we consider $m_\chi=1$ GeV, 25 GeV, and 40 GeV in the $Z$-mode, 
and $m_\chi=1$ GeV, 25 GeV, and 50 GeV in the $WW$-mode.}
\label{fig:mqzhist}
\end{centering}
\end{figure}

Fig.\ (\ref{fig:mqhist}) shows the 
normalized $E_\gamma$ and $\cos\theta_\gamma$ distributions 
for the signal  process \eexxa in the millicharged DM models, 
and for the irreducible background \eevva in the SM, 
in the CEPC $H$-mode with the detector cuts: 
$E_\gamma > 0.1$ GeV and $|\cos\theta_\gamma| < 0.99$.  
Fig.\ (\ref{fig:mqzhist}) shows the normalized $E_\gamma$ distributions
of the signal and background in the $Z$-mode and in the $WW$-mode, 
with the detector cuts: 
$E_\gamma > 0.1$ GeV and $|\cos\theta_\gamma| < 0.99$.


The millicharged DM models have different 
monophoton energy distributions for different DM 
masses. 
In all the three CEPC modes, 
the monophoton spectrum shows a peak 
structure towards the last energy bins, 
for the case where $m_\chi=1$ GeV. 
Whereas for higher DM masses, 
the monophoton spectrum steadily decreases 
with the increment of $E_\gamma$ 
and drops rapidly near the termination point 
$E_\chi^{m} = (s - 4 m_\chi^2)/(2 \sqrt{s})$.

\begin{figure}[htbp]
\begin{centering}
\includegraphics[width=0.4\columnwidth]{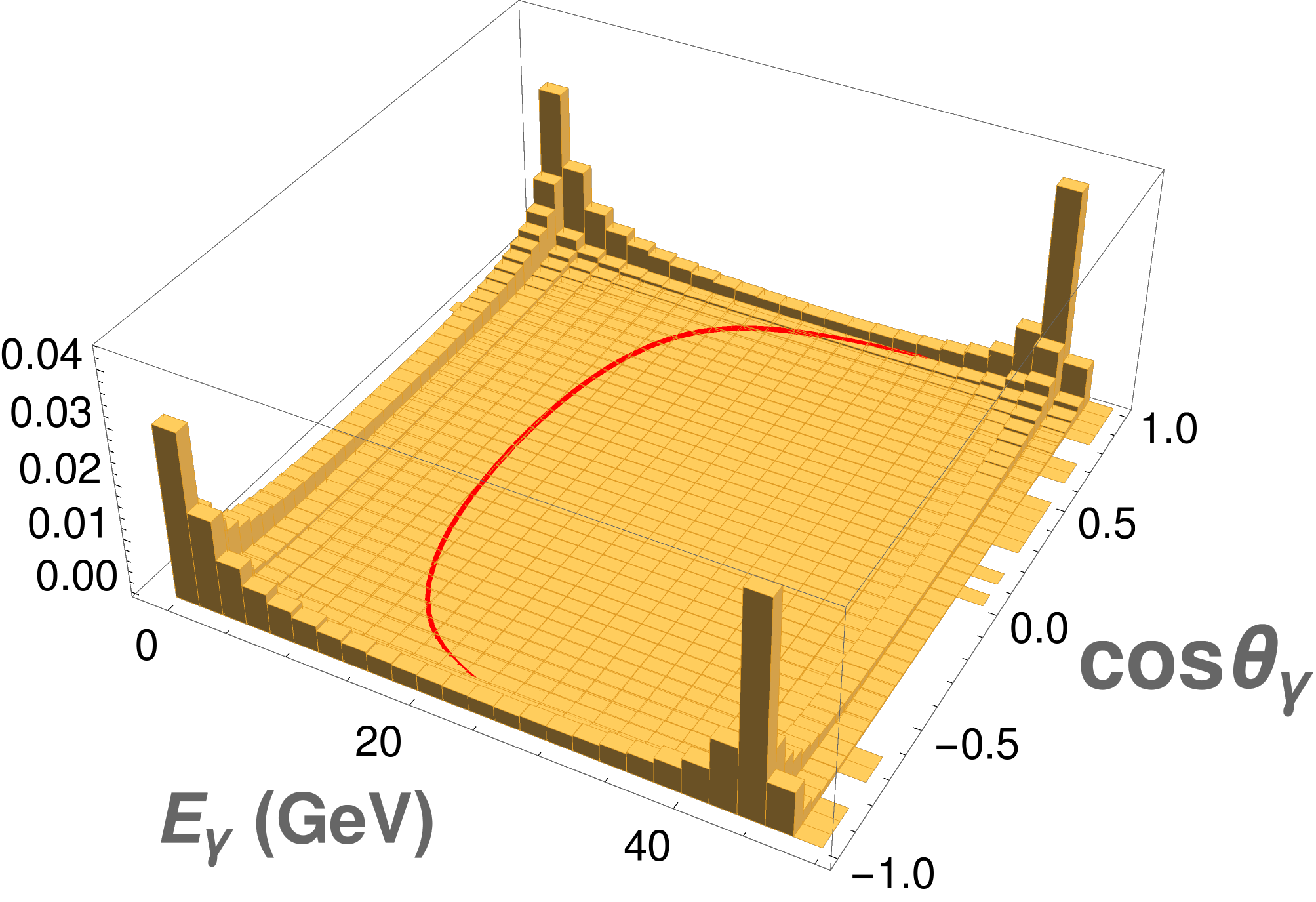}
\includegraphics[width=0.4\columnwidth]{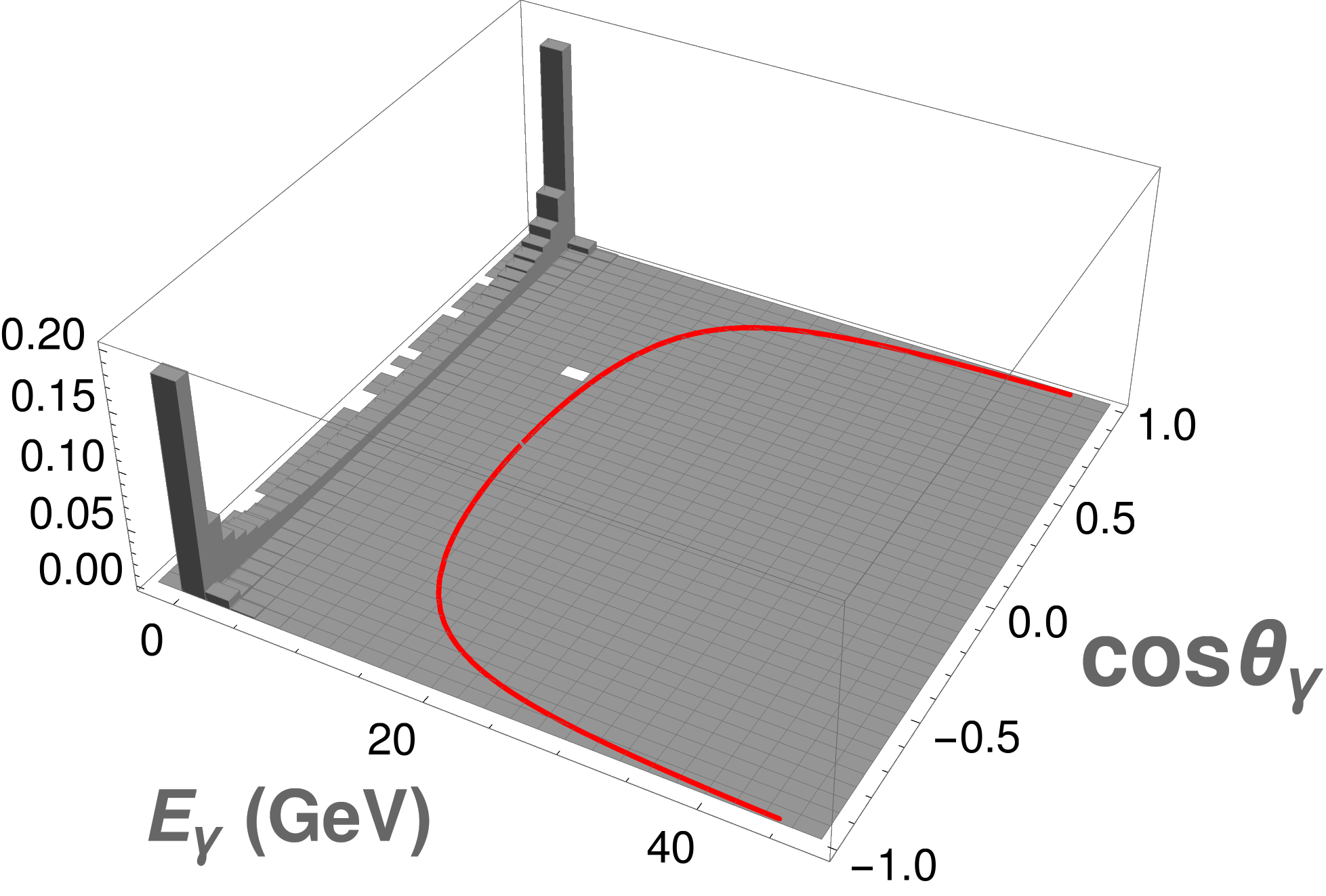}
\caption{Monophoton distributions in the 
$E_\gamma$-$\cos\theta_\gamma$ 
plane in millicharged DM models (left) 
and in SM irreducible background (right). 
Events on both plots are computed in 
the CEPC $Z$-mode. 
For the millicharged model, we use 
$m_\chi = 1$ GeV and $\varepsilon =0.01$. 
The red curves on both plots indicate  
$E_B^m(\theta_\gamma)$ 
as given in Eq.\ (\ref{eq:Em}). 
}
\label{fig:mqzhist3d}
\end{centering}
\end{figure}

The SM irreducible background exhibits 
different energy spectrum from the millicharged 
DM models. 
As shown in the left panel figure of Fig.\ (\ref{fig:mqhist}), 
the $Z$ resonance in the $H$-mode 
is located at  
$E^Z_\gamma \simeq 103$ GeV 
with $\Gamma^Z_\gamma \simeq 0.95$ GeV.
As shown in the right panel figure of Fig.\ (\ref{fig:mqzhist}), 
the $Z$ resonance in the $WW$-mode  
occurs at  
$E^Z_\gamma \simeq 54$ GeV 
with $\Gamma^Z_\gamma \simeq 1.4$ GeV. 
The $Z$ resonance in the monophoton spectrum both in  
the $H$-mode and in the $WW$-mode can be 
easily distinguished from the DM models. 
In the contrary, 
the $Z$ resonance in the $Z$-mode appears at 
$E^Z_\gamma \simeq 0$ GeV 
with $\Gamma^Z_\gamma \simeq 2.5$ GeV, 
which coincides with the low energy divergence 
behavior in the millicharged models, 
as shown in the left panel figure of Fig.\ (\ref{fig:mqzhist}). 
However, photon events with energy 
$E_\gamma \lesssim 0.12 \sqrt{s}$ 
are excluded in the analysis by the detector cut to remove 
the reducible SM background 
\footnote{In the $Z$-mode, the minimum 
energy cut for the reducible background 
$E_\gamma \gtrsim 0.12 \sqrt{s} \simeq 11$ GeV 
exceeds the maximum photon energy in the case 
where $m_\chi = 40$ GeV in the millicharged DM models. 
Thus, to optimize the 
sensitivity to millicharged DM models 
beyond $m_\chi \sim 40$ GeV, one has to relax the 
detector cut that is designed to eliminate the SM reducible 
background. We leave this to a future study.}. 
In the $Z$-mode, 
because the background events are highly focused in lower $E_\gamma$ region, 
as displayed in the left panel figure of Fig.\ (\ref{fig:mqzhist}) 
and in Fig.\ (\ref{fig:mqzhist3d}), 
the SM background can be suppressed more efficiently  
as compared to the other CEPC modes. 
Thus, in all the CEPC modes, 
removing the $Z$ resonance generally 
increases the capability of probing 
millicharged DM models.

In the analysis, we employ the basic detector cuts 
discussed in last section to remove 
the monophtoton events in the reducible background, 
the SM $Z$ resonance in the irreducible background, 
and the phase space where DM is unable to reach. 
Because the millicharged DM models exhibit similar angular distributions as the 
SM irreducible background in the monophoton channel, 
as shown in the right panel figure of Fig.\ (\ref{fig:mqhist}), 
we do not impose any further detector cut on 
$\theta_\gamma$ beyond the basic detector cuts. 
Because the monophoton spectrum exhibits a resonance 
feature near the end of the energy spectrum for the light DM masses, 
the signal significance can be enhanced in the $H$ and $WW$ modes 
if we only select the monophoton events on the right hand side of the $Z$ resonance. 
Thus, to improve the sensitivity in the $H$-mode ($WW$-mode),  
we impose the detector cut $E_\gamma > E_\gamma^Z(s) + 5 \Gamma_\gamma^Z(s)$ 
in addition to the basic detector cuts, 
for $m_\chi < 25$ (30) GeV. 
%

\begin{figure}[htbp]
\begin{centering}
\includegraphics[width=0.55\columnwidth]{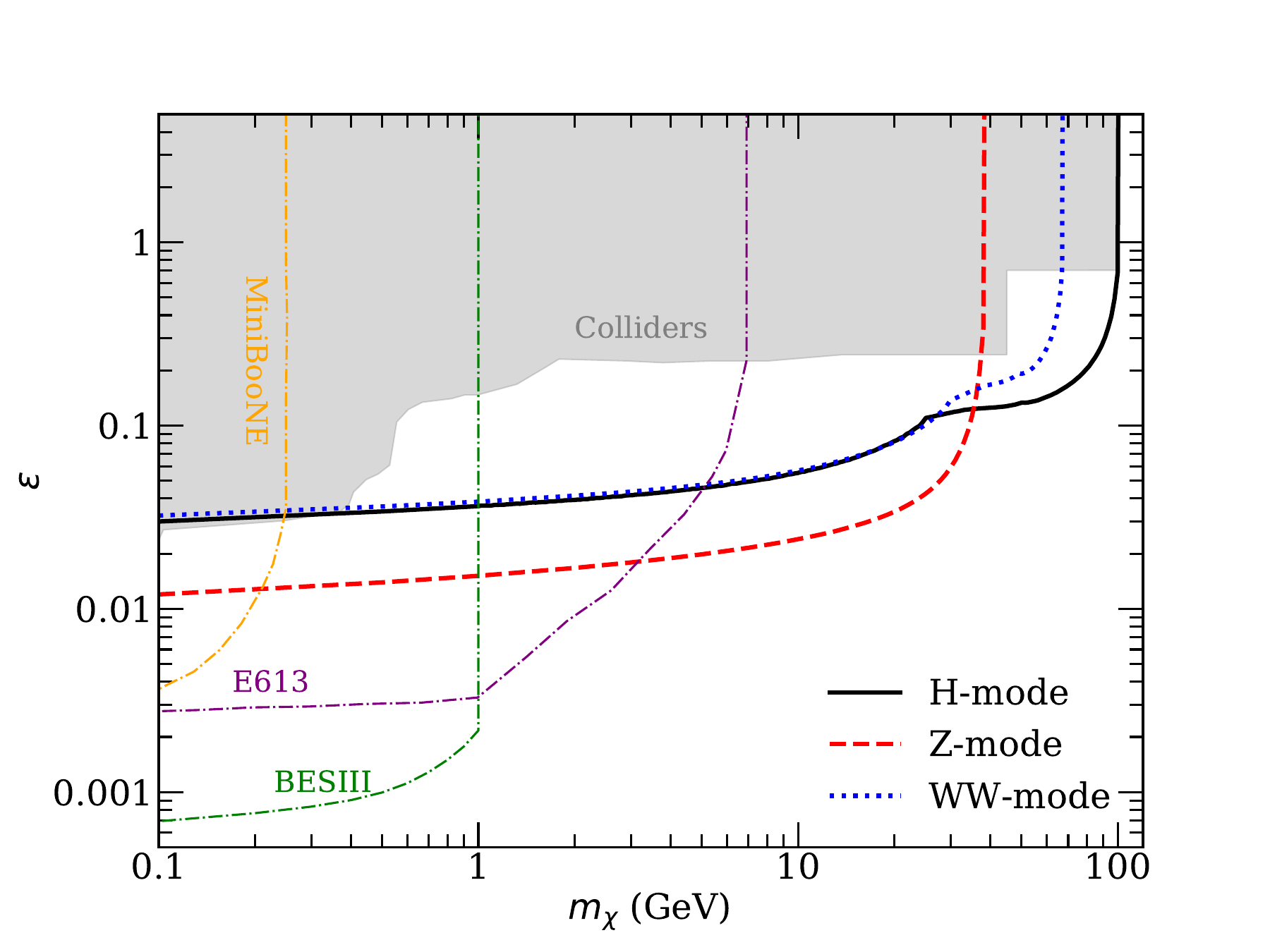}
\caption{Expected 95\% CL upper bounds on millicharge 
with 5.6 ab$^{-1}$ data in the $H$-mode (black-solid), 
with 16 ab$^{-1}$ data in the $WW$-mode (blue-dotted), 
and with 2.6 ab$^{-1}$ data in the $Z$-mode (red-dashed). 
Previous collider bounds are shown as 
the gray shaded region \cite{Davidson:2000hf}; 
expected sensitivity studied in previous analyses 
are also shown for comparison, including 
E613 \cite{Soper:2014ska}, 
MiniBooNE \cite{Magill:2018tbb}, 
and BESIII \cite{Liu:2018jdi}. 
}
\label{fig:mq-limit}
\end{centering}
\end{figure}

To study the 95\% C.L. upper bound on 
the production cross section in new physics models, 
and consequently on $\varepsilon$ in millicharged DM 
models, we use the simple criteria $N_S/\sqrt{N_B}=2$ 
to compute the reach throughout our analysis.
Fig.\ (\ref{fig:mq-limit}) shows the 95\% C.L. upper bound 
on millicharge $\varepsilon$ as the function of 
the DM mass $m_\chi$ in the millicharged DM model. 
Here we compute the limits based on 
5.6 ab$^{-1}$ data in the $H$-mode, 
16 ab$^{-1}$ data in the $WW$-mode, 
and 2.6 ab$^{-1}$ data in the $Z$-mode.  
The $Z$-mode has the best sensitivity 
for DM mass $\lesssim 40$ GeV, 
due to the fact that the production cross section 
in the millicharged DM model is larger 
and the SM irreducible background is smaller 
in the $Z$-mode than the other two modes. 
For the case where DM mass is 5 GeV, 
millicharge $\sim 0.02$ can be probed 
by the $Z$-mode running of CEPC. 
The $H$-mode has the best sensitivity 
for DM mass larger than 40 GeV. 
For the case where DM mass is $\sim 50$ GeV, 
millicharge $\sim 10^{-1}$ can be probed 
by the $H$-mode running of CEPC. 
The CEPC can probe a vast region of the 
parameter space that is previously unexplored, 
in the millicharged DM models 
for DM mass from 1 GeV to 100 GeV. 
Compared to previous collider bounds, 
the improvement on constraints on 
millicharge is about one order of magnitude 
for the mass from 1 GeV to 100 GeV.


The parameter space of millicharged DM to be probed in CEPC, 
$\varepsilon \lesssim (0.01-0.1)$ for (1-100) GeV mass, 
has been excluded by many orders of magnitude due to 
the interaction between the millicharged DM 
with the CMB \cite{McDermott:2010pa}. 
However, if only a small fraction of DM is millicharged 
(denoted as $f_{\rm mcp}$), the very 
stringent CMB constraints can be avoided
\cite{Dubovsky:2003yn} 
\cite{Cline:2012is}; 
the millicharged DM component would be  
misinterpreted as baryons in the current CMB data. 
Using Planck data, 
Ref.\ \cite{Dolgov:2013una} put 
an upper bound on the millicharged DM 
relic density, $\Omega_{\rm mcp}h^2 < 0.001$.  
Recently, EDGES 
experiment \cite{Bowman:2018yin} observed 
a stronger absorption signal than expected in the 21 cm data 
near redshift $z=17$; 
many works have studied whether one can use 
millicharged DM to explain such an anomaly  
while satisfying various constraints 
\cite{Munoz:2018pzp,Berlin:2018sjs, Barkana:2018cct, Slatyer:2018aqg, Kovetz:2018zan, Boddy:2018wzy,Wadekar:2019xnf}. 
In order to provide the sufficient cooling needed for the 21 cm anomaly, 
the millicharged DM particle should be in $\sim$ MeV scale 
assuming  $f_{\rm mcp} \lesssim 1\%$ \cite{Munoz:2018pzp}. 
Recent studies have further reduced the upper bound 
on the millicharged DM component: 
$f_{\rm mcp}<  0.6\%$ 
from helium fraction in the CMB \cite{dePutter:2018xte}; 
refs.\ \cite{Kovetz:2018zan, Boddy:2018wzy} found 
if $f_{\rm mcp}< 0.4\%$, the bound from current CMB data can 
be evaded.

Although astrophysical processes can yield 
very strong limits on millicharged DM, 
such analysis often relies on astrophysical 
assumptions of the millicharged particles. 
The collider constraints 
analyzed here with CEPC, 
however, 
do not make any astrophysical 
assumptions about the millicharged particles. 
Thus, the CEPC constraints given in this analysis for 
millicharged particles are usually more robust than many of 
those obtained in astrophysical processes.

	
\section{$Z'$ portal DM models}
\label{sec:zp}

We study in this section the potential constraints 
on the fermionic DM that interacts with the 
standard model sector via a $Z'$ portal. 
In our analysis, we select the model where 
$M_{Z'} = 150$ GeV as the benchmark point. 
The $Z'$ boson can interact with fermions via both 
vector and axial-vector couplings; 
for simplicity, in this study, only the situations in which 
$Z'$ couples to both SM fermions and DM  
in a pure vector manner ($g_A = 0$), 
or in a pure axial-vector manner ($g_V = 0$) 
are considered. 
To search for the collider signals  
from the $Z'$ portal DM model, 
we study the monophoton signature 
as well as visible decay from the $Z'$ boson, 
in this section. 

\begin{figure}[htbp]
\begin{centering}
\includegraphics[width=0.45\columnwidth]{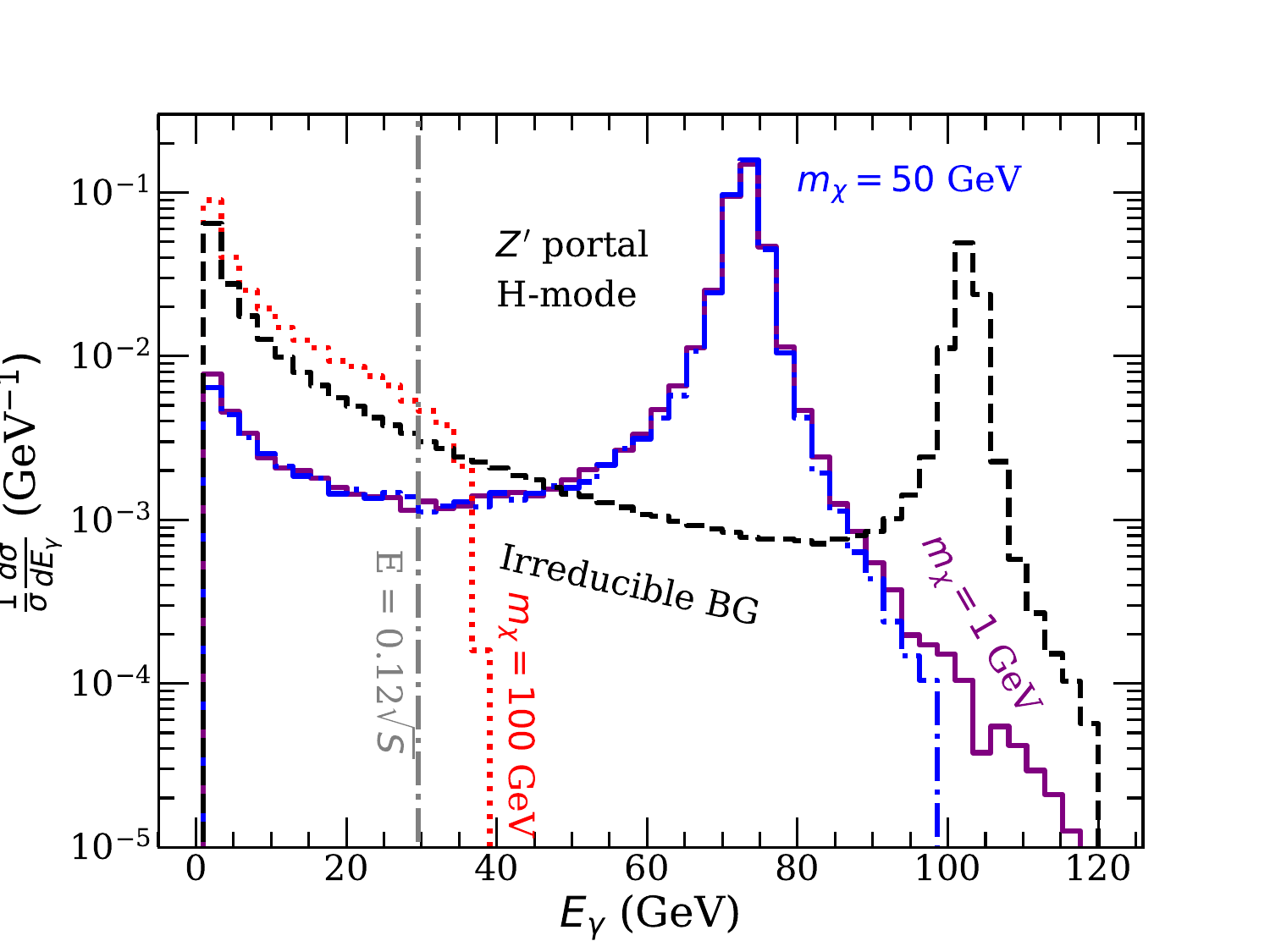}
\caption{Normalized $E_\gamma$ distribution 
at the CEPC $H$-mode, 
in the $Z'$ portal model, as well as 
in the SM irreducible background process.
%
Only the vector couplings are considered, 
i.e., $g^\chi_A=0$ and $g^f_A=0$. 
We use $g^\chi_V=1$, $g^f_V=0.01$, 
and $M_{Z^\prime}=150\GeV$ for 
the $Z'$ portal model.  
For the DM mass, we consider three 
different cases: 
$m_\chi=1$ GeV, 50 GeV, and 100 GeV.  
}
\label{fig: dfhhist}
\end{centering}
\end{figure}

The normalized monophoton energy distributions  
in the $Z'$ portal DM models where $M_{Z'}=150$ GeV 
are shown in Fig.\ (\ref{fig: dfhhist}), 
along with the SM irreducible background \eevva. 
Unlike the millicharged DM models, 
the monophoton spectrum 
in the $Z'$ portal DM models 
exhibits a peak at 
$E_\gamma \simeq (s-M_{Z'}^2)/(2\sqrt{s})$,
which is $\sim$ 73 GeV for the model considered here 
in the $H$-mode. 
Such a $Z'$ resonance is visible in Fig.\ (\ref{fig: dfhhist}) 
for the 
$m_\chi=1$ GeV and $m_\chi=50$ GeV cases; 
for the $m_\chi=100$ GeV case, however, 
the $Z'$ resonance can not appear 
because it exceeds the maximum photon energy 
in the final state, $E_\chi^{m}$, 
which is $\sim 37$ GeV here. 


Here we focus our analysis primarily on the $H$-mode. 
This is due to the fact that the 150 GeV $Z'$ can only be 
produced off-shell in the $Z$-mode. 
Although the 150 GeV $Z'$ can be produced on-shell 
in the $WW$-mode, 
the final state photons from the on-shell $Z'$ have  
$E_\gamma \sim$ 10 GeV which falls below the detector 
cut for the reducible background $0.12\sqrt{s}$. 
Thus for the $Z$ mode, we only apply the 
basic detector cuts; 
for the $WW$ and $H$, additional detector 
cuts are applied to maximize the CEPC 
sensitivity.  
For the $WW$ mode, we further veto events 
with $E_\gamma > E_\gamma^Z-5\Gamma_Z$ 
on top of the basic detector cuts. 
For the $H$ mode, we always select events in 
the $Z'$ resonance if it is present; 
thus for $m_\chi \leq 75 \GeV$ in the $H$ mode, 
we require $147 \GeV < M_\gamma < 153 \GeV$ 
where  $M_\gamma = \sqrt{s-2 \sqrt{s} E_\gamma}$.  


We can also search for the $Z'$ boson via its 
visible decay channels at CEPC 
(see e.g., \cite{Karliner:2015tga} \cite{He:2017zzr} 
for previous studies on this topic). 
We take the $\mu^+\mu^-$ final state as the 
visible channel to probe the $Z'$ portal DM models 
in this section. 
We adopt the muon momentum resolution  
as follows 
\be
{\delta p_T \over p_T} = {p_T \over 10^5\GeV} 
\bigoplus 0.1\% ~{\rm for}~  |\eta| < 1.0
\label{eq:mu}
\ee
and 10 times greater for $1.0<|\eta|<3.0$, 
which has been implemented in the 
CEPC card in Delphes \cite{deFavereau:2013fsa}. 
The muon momentum resolution is 
much better than the photon energy resolution, 
as given in Eq.\ (\ref{eq:resolution}). 

\begin{figure}[htbp]
\begin{centering}
\includegraphics[width=0.45\columnwidth]{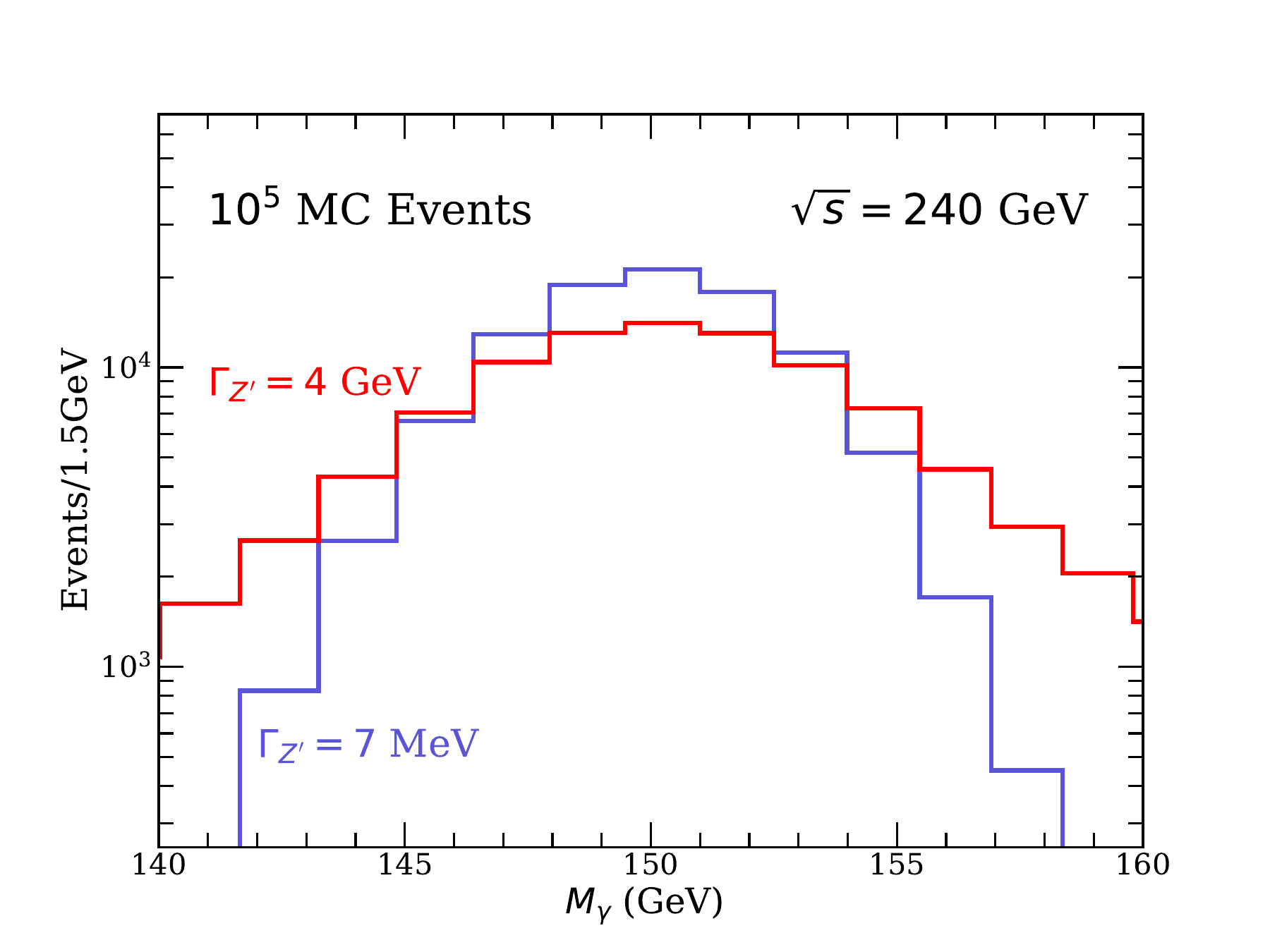}
\includegraphics[width=0.45\columnwidth]{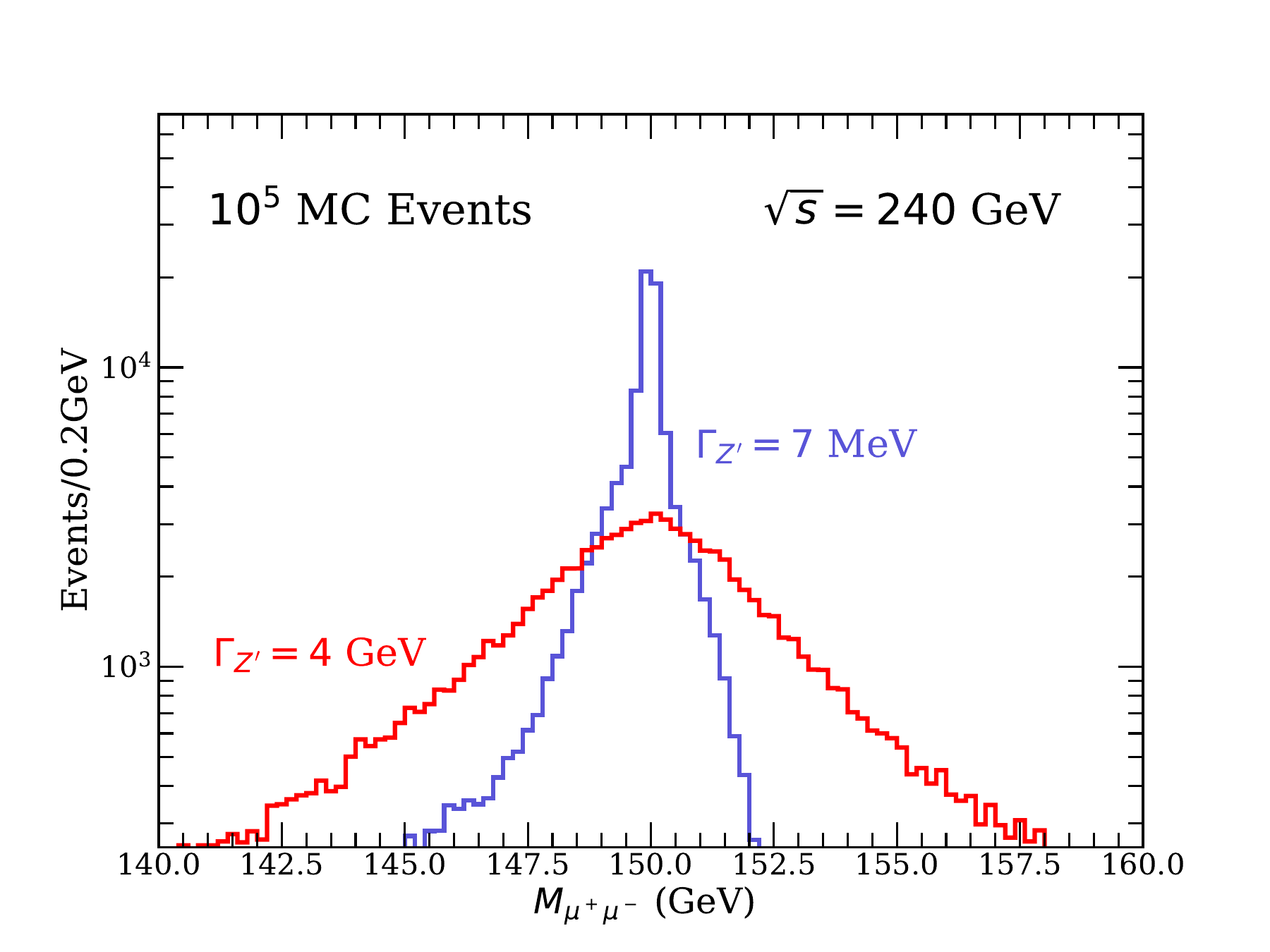}
\caption{Left panel: $M_\gamma$ distribution 
in $e^+e^-\to \gamma Z^\prime \to \bar{\chi}\chi\gamma$ 
in the CEPC $H$-mode, 
where $M_\gamma = \sqrt{s-2 \sqrt{s} E_\gamma}$.
Right panel: $M_{\mu^+ \mu^-}$ distribution 
in $e^+e^-\to \gamma Z^\prime \to \mu^+ \mu^- \gamma$ 
in the CEPC $H$-mode. 
We consider the vector coupling case (i.e., $g_A=0$), 
where $g^\chi_V =1$ and $g^f_V = 0.01$. 
The $Z'$ boson mass is fixed at $M_{Z'} = 150$ GeV. 
We consider two different DM masses: 
(1) $m_{\chi} = 0$ GeV so that the $Z'$ boson has 
a large decay width $\Gamma_{Z'} \simeq 4.0$ GeV; 
(2) $m_{\chi} = 75$ GeV so that the $Z'$ boson 
has a narrow width $\Gamma_{Z'} \simeq 7.0$ MeV.  
Simulations in both channels are carried out 
with MadGraph and Delphes. 
}
\label{fig:visible}
\end{centering}
\end{figure}

We reconstruct the $Z'$ boson resonance in the monophoton 
channel as well as in the  $\mu^+ \mu^-$ channel, in Fig.\ (\ref{fig:visible}). 
The detector simulations are carried out 
in MadGraph \cite{Alwall:2014hca}, 
Pythia 8 \cite{Sjostrand:2007gs}, 
and Delphes 3 \cite{deFavereau:2013fsa} 
for both channels. 
For the monophoton channel, the $Z'$ resonance is reconstructed via 
$M_\gamma = \sqrt{s-2 \sqrt{s} E_\gamma}$; 
for the dimuon channel, the $Z'$ resonance is reconstructed via 
the invariant mass of the muon pair. 
The mass resolution in the monophoton channel can be 
computed as 
$ \delta M_\gamma = (\sqrt{s}/M_\gamma)\delta E_\gamma$ 
which is about 1.5 GeV for $M_\gamma=150$ GeV. 
For the $Z'$ resonance in the di-muon invariant mass 
reconstruction, we apply the resolution given in 
Eq.\ (\ref{eq:mu}) for $p_T\simeq 50$ GeV to 
energy, transverse momentum, and longitudinal 
momentum for each muon, 
and estimate the mass resolution to 
be $\simeq 0.2$ GeV for $M_{\mu^+\mu^-}=150$ GeV. 
We thus use 1.5 (0.2) GeV as the width in the mass reconstruction  
to bin the events from the monophoton (dimuon) channel, 
in Fig.\ (\ref{fig:visible}). 

Two benchmark models are shown in Fig.\ (\ref{fig:visible}) 
where the $Z'$ boson has a decay width 
$\sim$ 4 GeV (7 MeV) 
for $m_\chi = 0$ (75) GeV. 
For the $m_\chi = 0$ case, 
the resonances reconstructed via both channels exhibit 
the $Z'$ intrinsic width. 
However, for the $m_\chi = 75$ GeV case, 
because the intrinsic $Z'$ decay width is smaller than the 
resolution in the photon channel as well as in 
the di-muon channel, 
the reconstructed $Z'$ peaks in both channels 
exhibit the widths due to detector resolutions. 
Thus, we select dimuon events in the invariant 
mass window $M_{\mu\mu} \in (150 \pm 3)$ GeV 
for $m_\chi<75$ GeV, 
and 
$M_{\mu\mu} \in (150 \pm 0.5)$ GeV 
for $m_\chi>75$ GeV. 

%


In our $Z'$ portal DM model, we assume that 
the $Z'$ boson has a universal coupling to both 
charged leptons and to quarks. 
Thus, one expects a recoil signal 
arising from the $Z'$ portal DM model 
in dark matter direct detection experiments that look for weakly 
interacting massive particles (WIMPs). 
Below we consider two different cases. 
First, we consider the case where the $Z'$ boson 
couples both to SM fermions and to DM fermion via 
vector couplings; in this case, the dark matter 
direct detection cross section is dominated 
by the spin-independent 
(SI) cross section which is given by 
\be
\sigma^{\rm SI}_{n\chi} = \sigma^{\rm SI}_{p\chi} 
= {9 \over \pi}
{(g_V^fg_V^\chi \mu_{n\chi})^2 \over M_{Z^\prime}^4}
\ee
where $\mu_{n\chi}$ is the reduced mass of 
the DM and nucleon. 
We also consider the case where the $Z'$ boson 
couples both to SM fermions and to DM fermion via 
axial-vector couplings; in this case, the dark matter 
direct detection cross section is dominated 
by the spin-dependent 
(SD) cross section which is given by 
\cite{Aprile:2019dbj}
\cite{Boveia:2016mrp} 
\be 
\sigma^{\rm SD}_{n\chi} = \sigma^{\rm SD}_{p\chi} 
\simeq {0.31 \over \pi}
{(g_A^fg_A^\chi\mu_{n\chi})^2 \over M_{Z^\prime}^4}.
\ee


\begin{figure}[htbp]
\begin{centering}
\includegraphics[width=0.45\columnwidth]{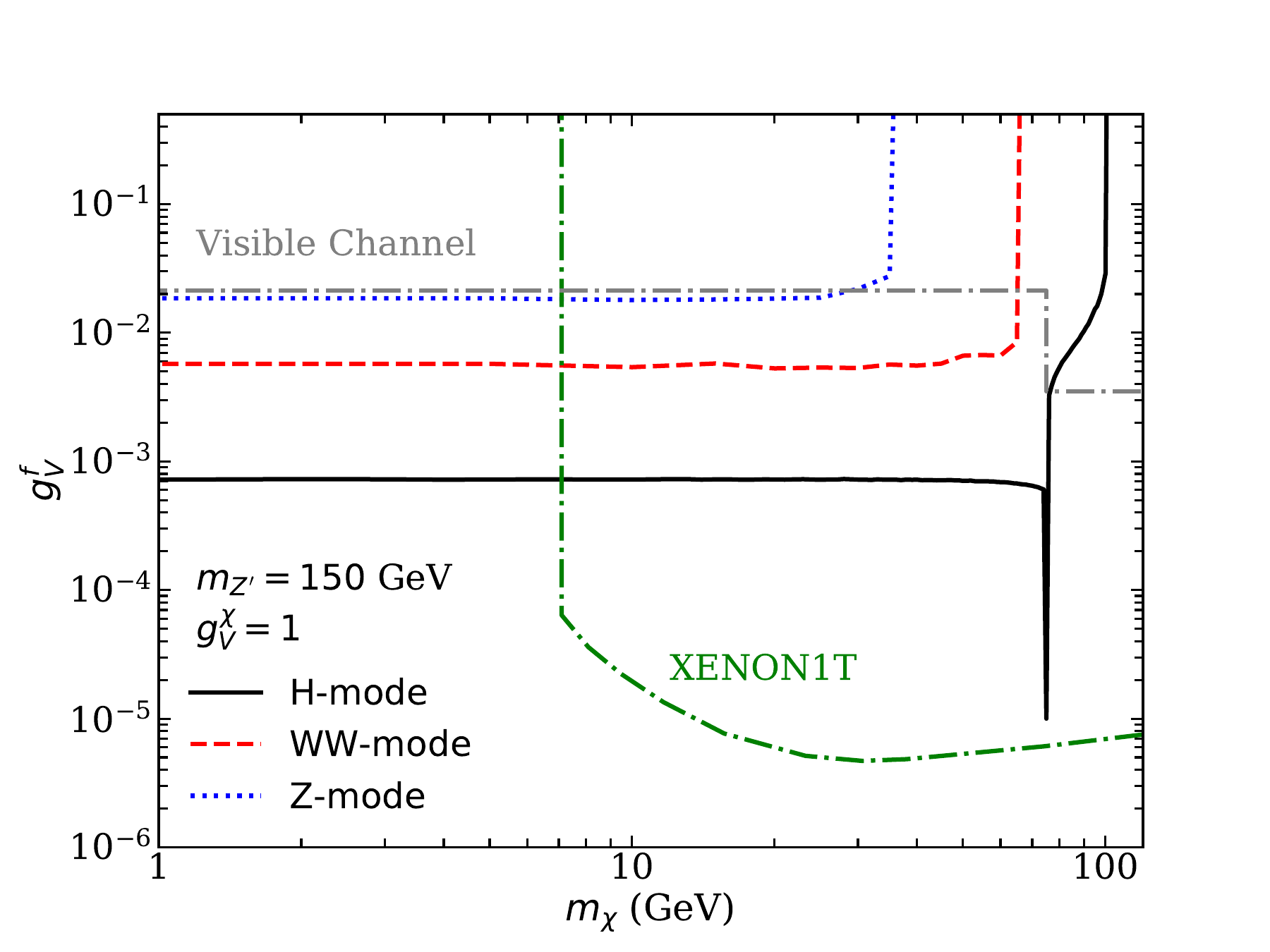}
\includegraphics[width=0.45\columnwidth]{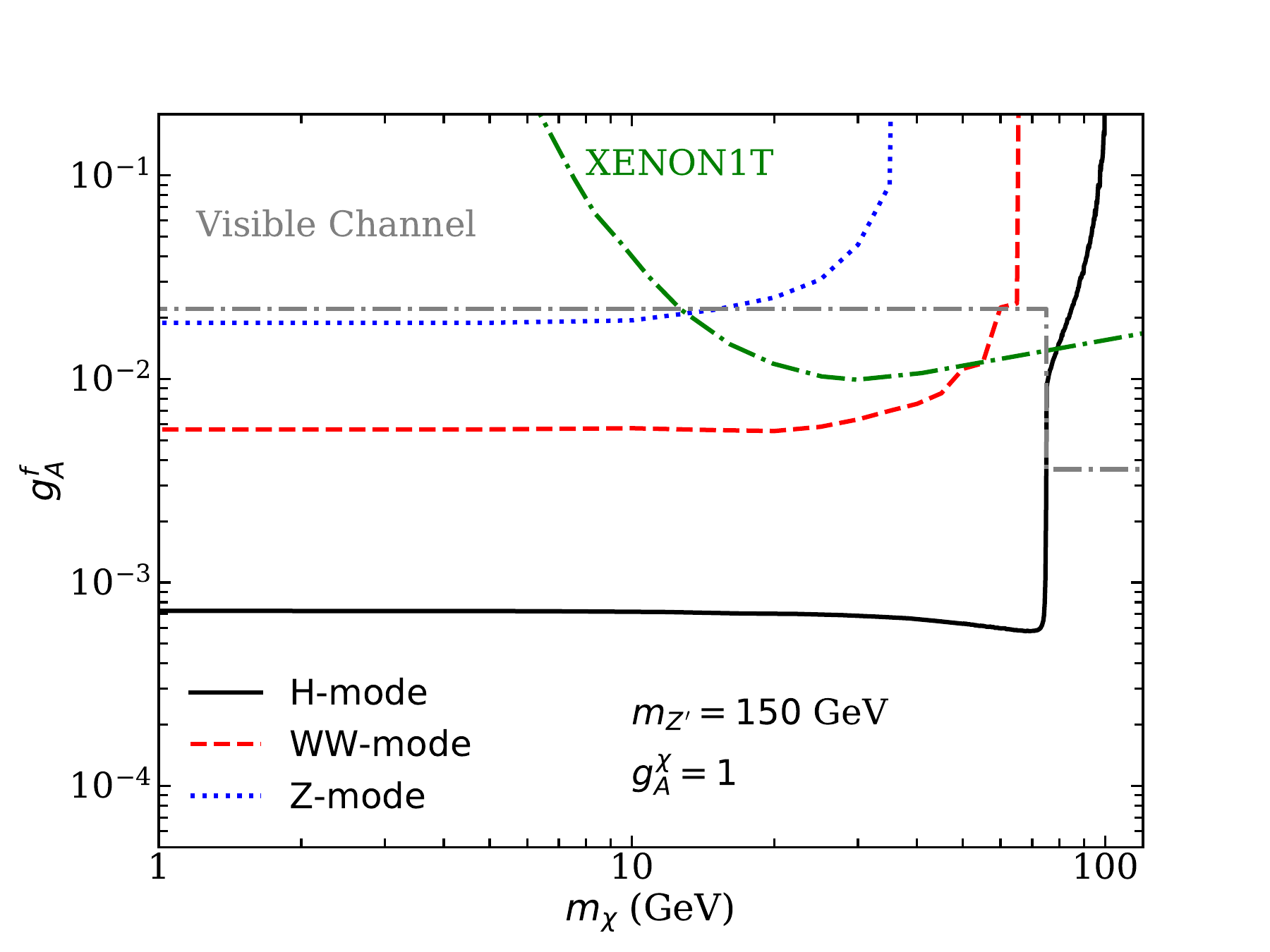}
\caption{Expected CEPC limits on $g^f_V$ (left), 
and on $g^f_A$ (right), 
in the $Z^\prime$ portal DM model where $M_{Z'} =150$ GeV, 
from the monophoton channel 
$e^+e^-\to\gamma Z'\to\gamma\bar{\chi}\chi$ 
in the three CEPC running modes. 
On the left (right) panel plot, 
only vector (axial-vector) couplings are assumed, 
with $g_V^\chi=1$ (left) and $g_A^\chi=1$ (right). 
Also shown are limits from 
$e^+e^-\to\gamma Z'\to\gamma \mu^+ \mu^-$ in the $H$-mode, %
from SI limit \cite{Aprile:2018dbl} (left) 
and SD limit \cite{Aprile:2019dbj} (right) in Xenon1T. 
}
\label{fig:zpRes}
\end{centering}
\end{figure}

In our analysis, we assume that the $Z'$ boson interacts with DM 
via an ${\cal O}(1)$ coupling, and has a rather weak 
coupling strength with the SM sector. 
We consider two cases where $Z'$ couples with fermions 
via only vector couplings, and via only axial-vector couplings. 
The left panel figure of Fig.\ (\ref{fig:zpRes}) shows the 
95\% CL upper bound on $g_V^f$ where $g_V^\chi=1$, 
in the vector coupling only case; 
the right panel figure of Fig.\ (\ref{fig:zpRes}) shows the 
95\% CL upper bound on $g_A^f$ where $g_A^\chi=1$, 
in the axial-vector coupling only case. 
For light dark matter mass, the monophoton channel in the 
$H$-mode has the best sensitivity in 
probing the gauge couplings between the 
150 GeV $Z'$ boson and the SM fermions
for both vector and axial-vector cases, 
as shown in Fig.\ (\ref{fig:zpRes}). 
CEPC can probe both $g_V^f$ and $g_A^f$ down 
to $\sim 7\times 10^{-4}$ 
for $m_\chi < M_{Z'}/2 = 75\GeV$ via 
monophoton searches. 
However, for $m_\chi > 75\GeV$, 
the visible decays of the 
$Z'$ boson become the better channel to 
study the $Z'$ model considered 
where the monophoton channel quickly loses 
its sensitivity, 
as shown on both plots in Fig.\ (\ref{fig:zpRes}). 
Interestingly, there is a sudden increase in the 
sensitivity in the monophoton channel when 
the dark matter mass approaches  
75 GeV from below so that a very 
narrow dip structure is shown near $m_\chi=75$ GeV, 
in the vector coupling only case. 
There is also an increased sensitivity in the di-muon channel 
when $m_\chi$ becomes larger than 75 GeV, due to the 
change of the detector cuts as the $Z'$ width turns narrower 
for $m_\chi$ crosses 75 GeV. 
We also compute the upper bound from Xenon1T 
experiment, including 
the SI limit \cite{Aprile:2018dbl} for the vector coupling case, 
and the SD limit \cite{Aprile:2019dbj} for the axial-vector coupling case. 
The Xenon1T limit is stronger than the CEPC limit 
for $m_\chi \geq 5$ GeV in the vector only case. 
For the axial-vector only case, however, the CEPC 
limits are typically better than the current Xenon1T limits. 
We apply Xenon1T limits to the $Z'$ portal DM models 
based on the assumption that the $Z'$ boson couples 
to both charged leptons and quarks with equal coupling 
strength. If the $Z'$ boson only interacts with electrons in 
the SM sector, the Xenon1T limits analyzed is no longer 
applicable. 

\begin{figure}[htbp]
\begin{centering}
\includegraphics[width=0.4\columnwidth]{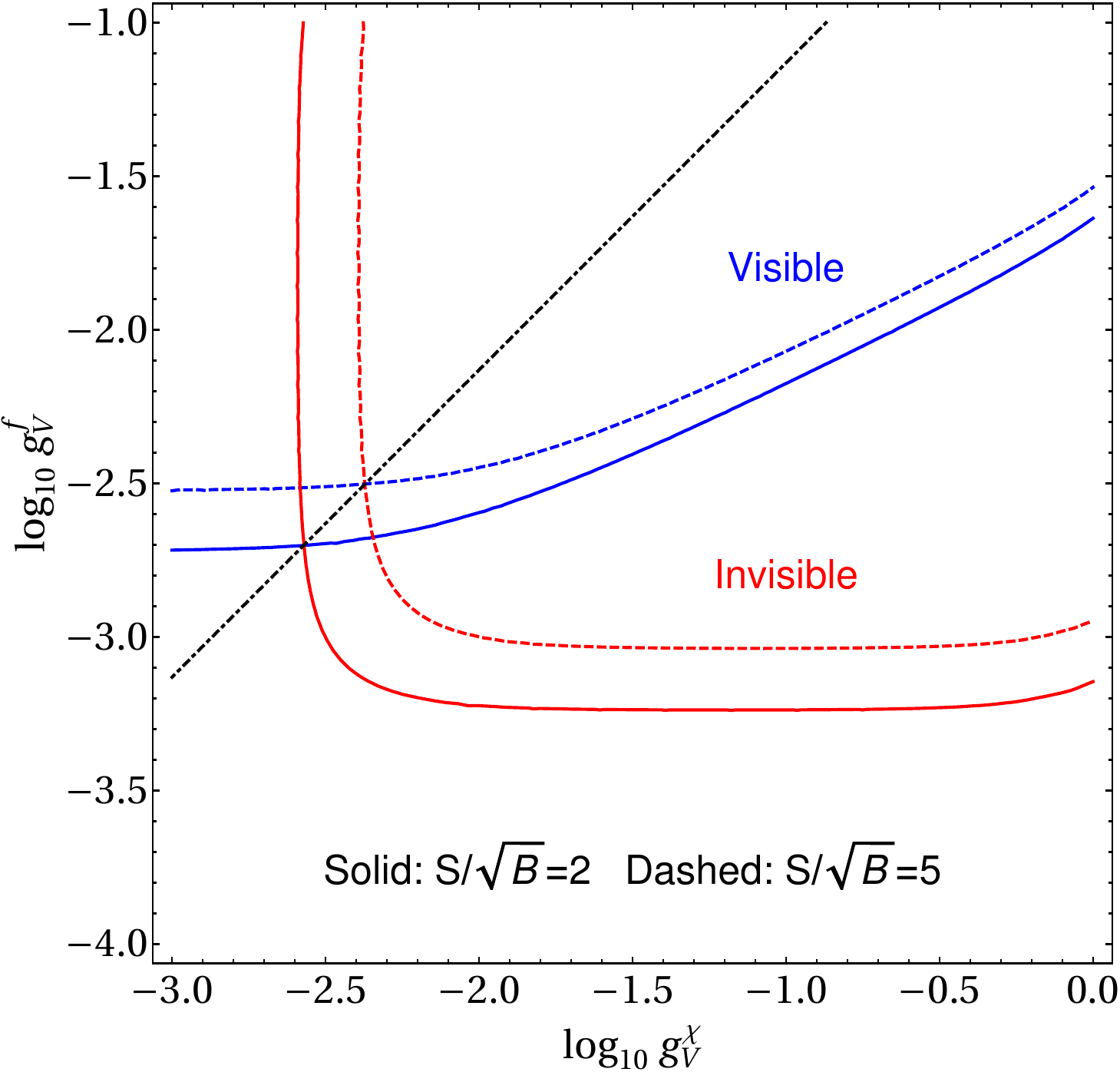}
\caption{The CEPC sensitivity on the $g_V^\chi$-$g_V^f$ plane 
in the $H$-mode 
via the monophoton channel $e^+ e^- \to \gamma Z' \to \gamma \chi \chi$ 
(the invisible channel), 
and via the dimuon channel 
$e^+ e^- \to \gamma Z' \to \gamma \mu^+ \mu^-$ 
(the visible channel). 
We consider the vector coupling only case 
with $M_{Z^\prime} = 150$ GeV and $m_\chi = 50$ GeV. 
The upper bounds in the visible (invisible) channel 
are shown in blue (red) lines assuming 5.6 ab$^{-1}$ data 
in the $H$-mode; 
the solid (dashed) lines indicate 
the 2 $\sigma$ (5 $\sigma$) reaches. 
The black dot-dashed line divides the parameter space 
into two regions: the invisible (visible) channel is the better 
channel in the region below (above) the black line. 
Here only di-muon events in the invariant mass window 
$150\pm 3$ GeV are selected.
}
\label{fig:dfexclusion_couplings}
\end{centering}
\end{figure}

We further study the constraints from the 
monophoton channel and from the di-muon 
channel to the two dimensional parameter 
space spanned by $g_V^\chi$ and $g_V^f$ 
in the vector coupling case where 
$M_{Z^\prime} = 150$ GeV 
and $m_\chi = 50$ GeV, 
as shown in Fig.\ (\ref{fig:dfexclusion_couplings}). 
The monophoton channel (invisible) 
usually provides a better constraint than 
the di-muon channel (the visible channel) 
in the parameter space $g_V^\chi \gtrsim g_V^f$, 
and vice versa. 


\section{DM effective operators}
\label{sec:eft}

\begin{figure}[htbp]
\begin{centering}
\includegraphics[width=0.45\columnwidth]{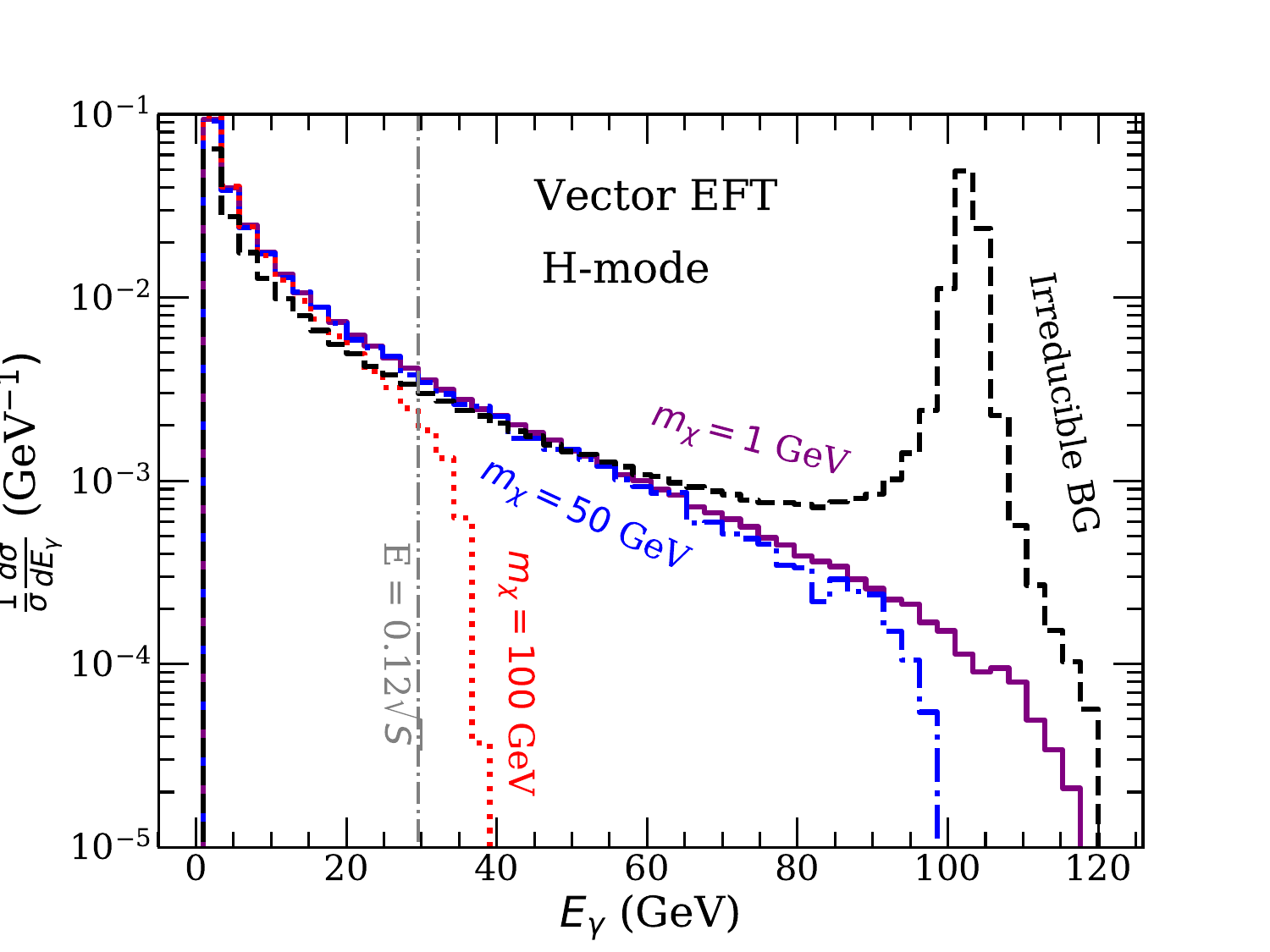}
\caption{Normalized $E_\gamma$ distribution in \eexxa 
for DM effective operator (the vector case), 
along with the SM irreducible background, 
in the CEPC $H$-mode.}
\label{fig:eftdis}
\end{centering}
\end{figure}

In this section, we study the potential CEPC constraints 
on DM effective operators.
We consider the four effective operators 
as given in Eq.\ (\ref{eq:eftl}). 
Fig.\ (\ref{fig:eftdis}) presents the normalized $E_\gamma$ distribution
at the CEPC in the $H$-mode for the process \eexxa
in the vector operator case, 
along with the irreducible background \eevva
in the SM. 
Unlike the millicharged DM models and the $Z^\prime$ portal 
DM models, the monophoton energy 
spectrum in the vector operator case 
does not exhibit any peak structure 
in the hard photon region. 
Thus we use the basic detector cuts 
for the DM effective operator analysis in this section; 
for the $H$-mode, we further veto photon events with 
$E_\gamma > 95 \GeV$ 
in addition to the basic detector cuts, 
to improve the CEPC sensitivity. 


\begin{figure}[htbp]
\begin{centering}
\includegraphics[width=0.4\columnwidth]{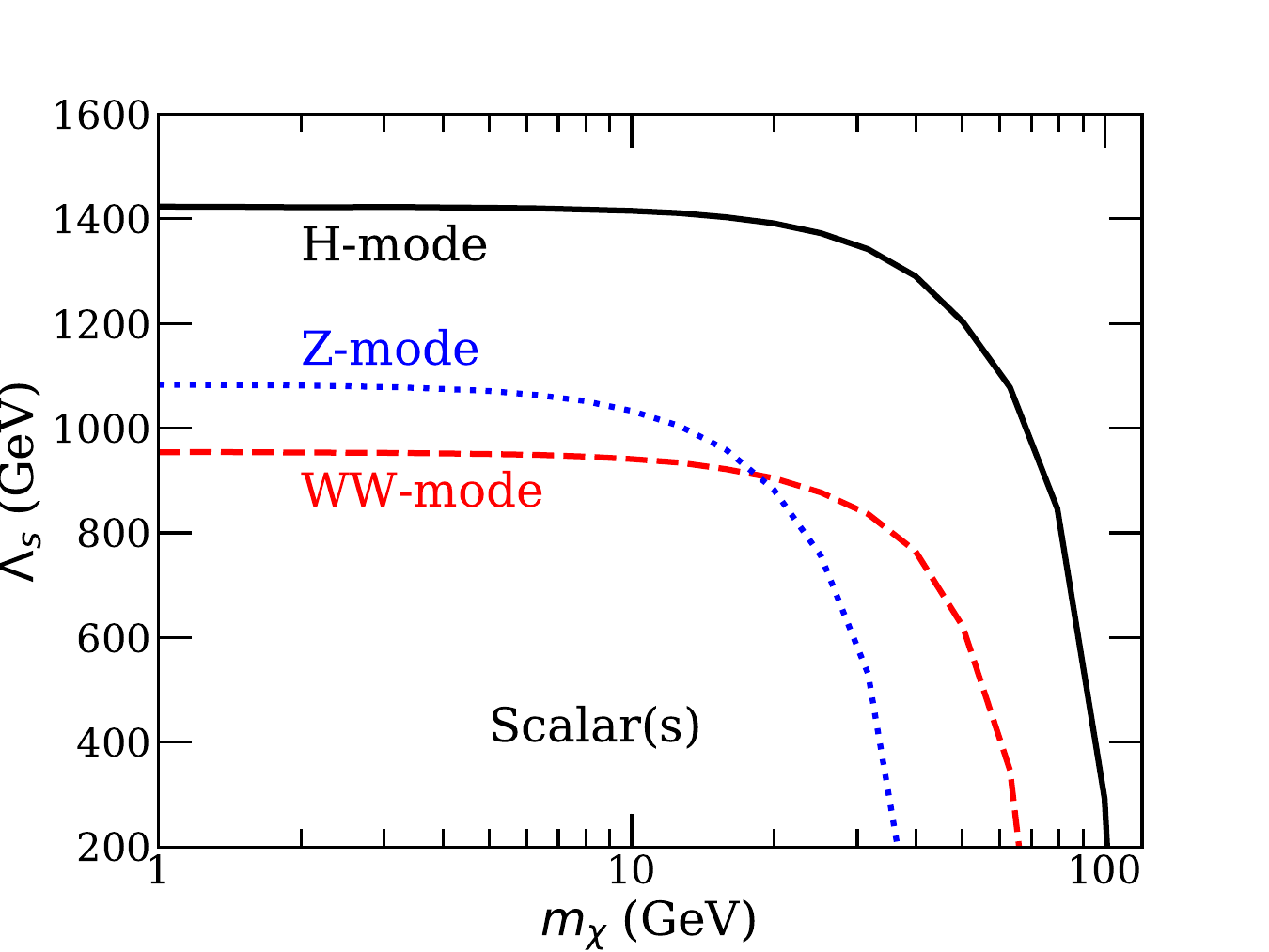}
\includegraphics[width=0.4\columnwidth]{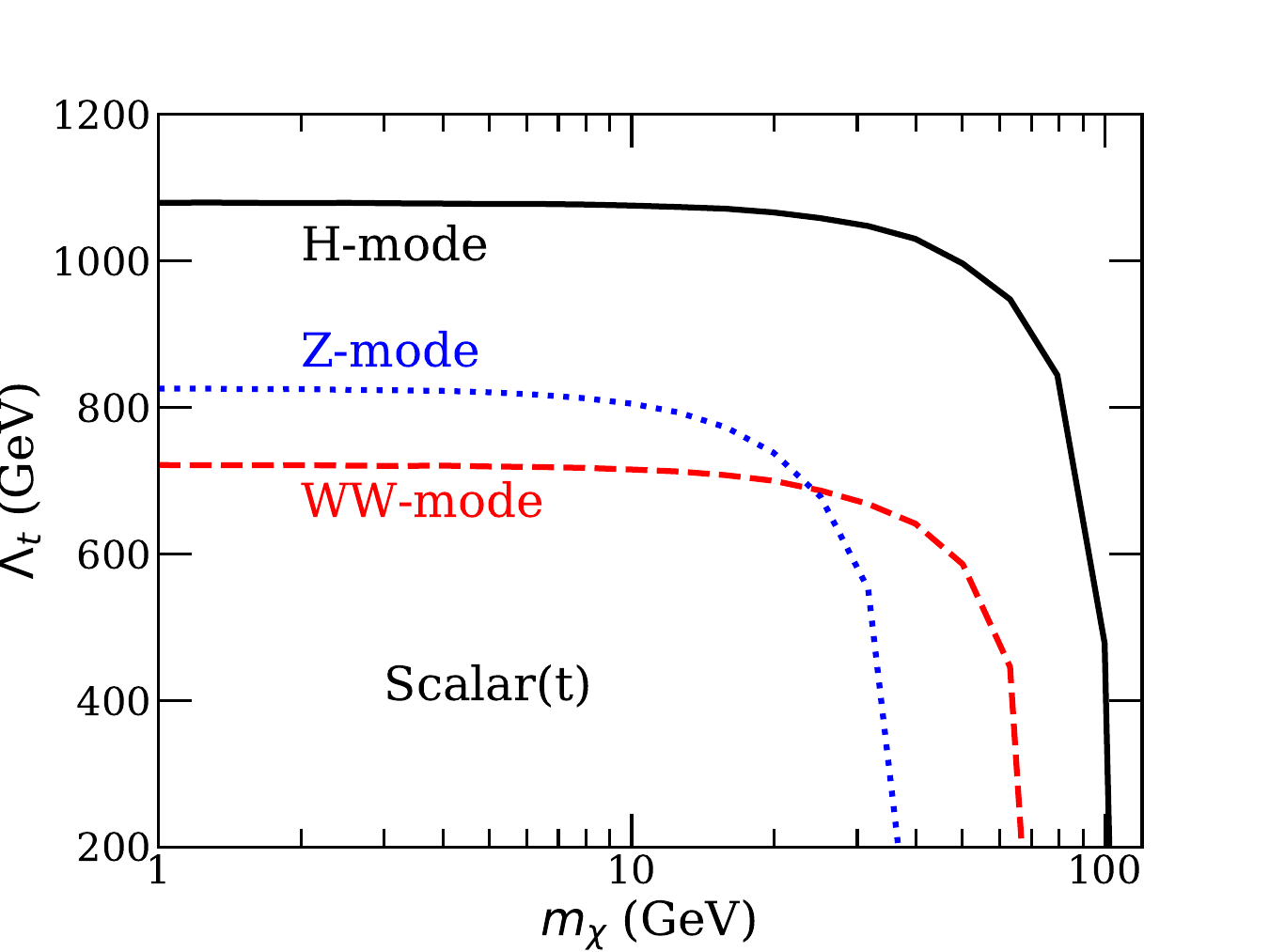}
\includegraphics[width=0.4\columnwidth]{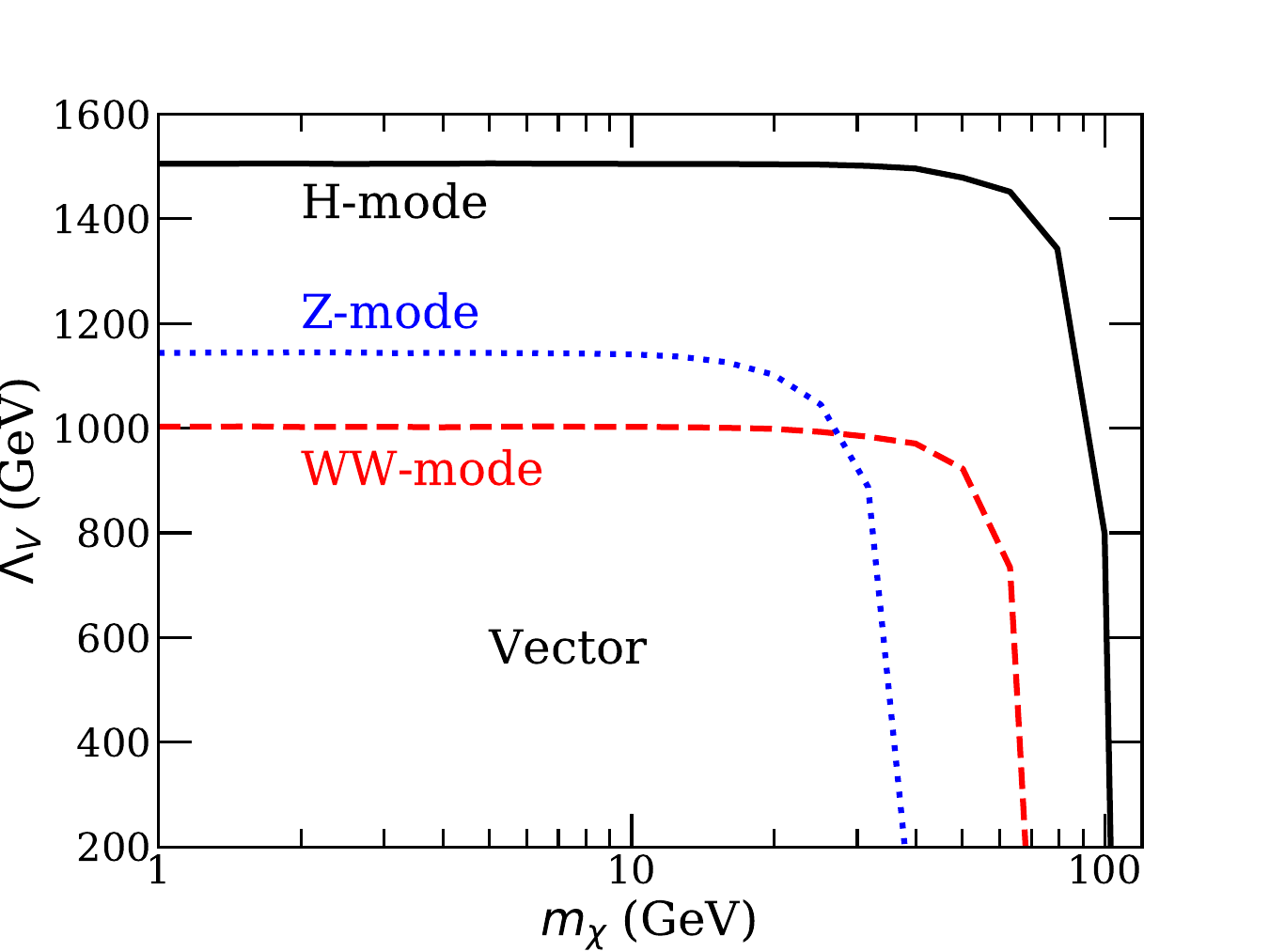}
\includegraphics[width=0.4\columnwidth]{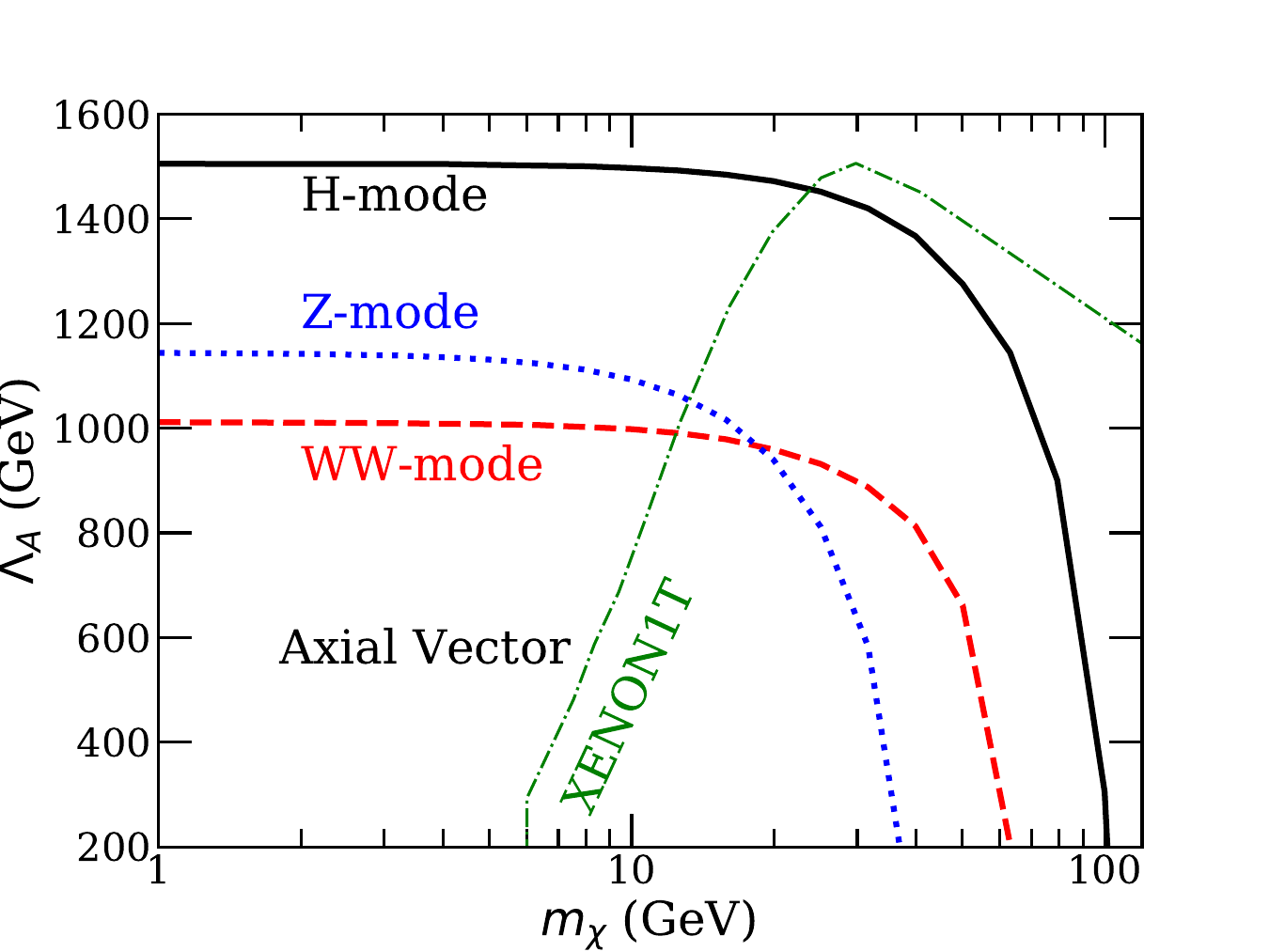}
\caption{Expected CEPC 95\% CL lower bound on 
$\Lambda$ with four different 
DM effective operators: 
the s-channel scalar interaction (upper left), 
the t-channel scalar interaction (upper right), 
the vector interaction (lower left), 
the axial-vector interaction (lower right). 
The SD limit from Xenon1T 
\cite{Aprile:2019dbj} 
is drawn for the axial-vector operator case, 
under the assumption that 
the effective operator coupling is universal.}
\label{fig:eft}
\end{centering}
\end{figure}

Fig.\ (\ref{fig:eft}) shows the 95\% C.L. lower
bound on the new physics characteristic scale $\Lambda$
as the function of the DM mass $m_\chi$ 
for the four DM effective operators considered. 
We analyzed all the three running modes at the CEPC with 
the integrated luminosity as 5.6 ab$^{-1}$ in the $H$-mode,
2.6 ab$^{-1}$ in the $WW$-mode, and 16 ab$^{-1}$ in the $Z$-mode. 
For all the four DM effective operators considered,
the CEPC $H$-mode has the best sensitivity to 
$\Lambda$ owing to the large production cross section; 
the $Z$-mode outperforms the $WW$-mode due to 
the larger integrated luminosity.

\begin{figure}[htbp]
\begin{centering}
\includegraphics[width=0.45\columnwidth]{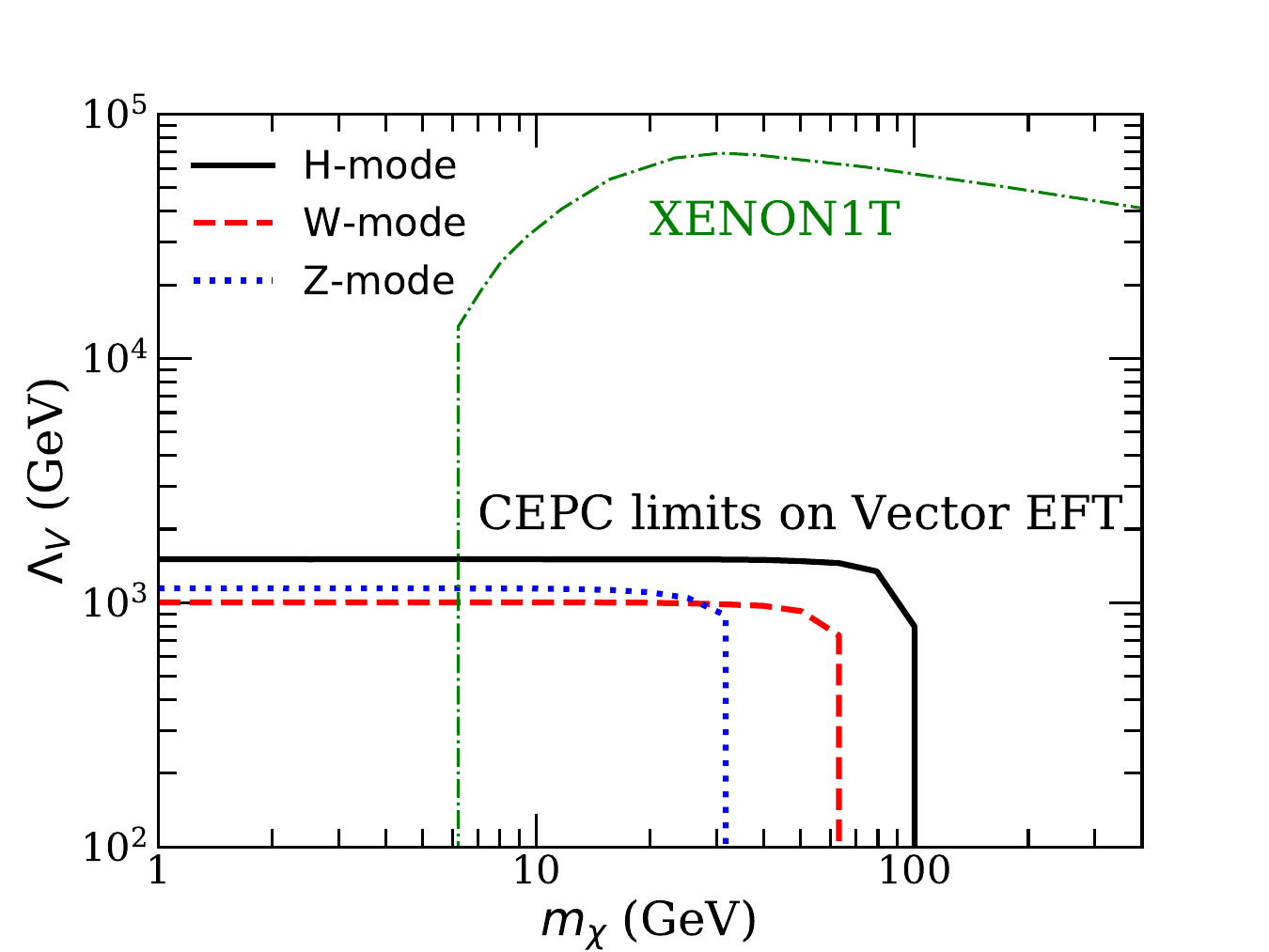}
\caption{Comparison between CEPC sensitivity and SI limits 
from Xenon1T
\cite{Aprile:2018dbl} 
under the assumption that the EFT couplings 
are universal for SM fermions.}  
\label{fig:eftSI}
\end{centering}
\end{figure}

We further compute the dark matter detection limits on 
DM effective operators, under the assumption that DM 
couples to all SM fermions with universal couplings. 
For the vector, s-scalar, and t-scalar effective operators, 
the dominant contribution to the dark matter direct 
detection experiments is the SI cross section; 
for the axial-vector effective operator case, 
the dominant contribution 
to the dark matter direct detection experiments 
is the SD cross section. 
For the SD limit, we use 
\begin{equation}
\sigma^{\rm SD}_{n\chi} = \sigma^{\rm SD}_{p\chi} 
= \frac{0.31}{\pi}\frac{\mu_{n\chi}^2}{\Lambda^4}; 
\end{equation}
for the SI limit, we use 
\begin{equation}
\sigma^{\rm SI}_{n\chi} = \sigma^{\rm SI}_{p\chi} 
= \frac{9}{\pi}\frac{\mu_{n\chi}^2}{\Lambda^4}
\end{equation}
where we have assumed that $\Lambda$ 
takes the same value for all charged lepton flavors and 
for all quark flavors. 
The lower bound on $\Lambda$ from the SD limit 
in Xenon1T experiment 
\cite{Aprile:2019dbj} 
is given in the lower-right panel 
plot in Fig.\ (\ref{fig:eft}), for the axial-vector {case}. 
The lower bound on $\Lambda$ from the SI limit 
in Xenon1T experiment 
\cite{Aprile:2018dbl} 
is given in Fig.\ (\ref{fig:eftSI}), 
for the vector operator case. 
The constraints from SI limits on the two scalar operator 
cases are similar to the vector operator case. 
We note that the Xenon1T constraints presented here 
are no longer valid if DM couples only with electron via 
effective operators.

\section{Detector cut optimization}
\label{sec:opt}


We present a preliminary study on optimizing 
the detector cuts to enhance the CEPC capability 
in probing the parameter space of the dark matter models 
considered. 
To do so, 
we first divide the two dimensional signature space spanned by 
$E_\gamma$ and $\cos\theta_\gamma$ into 900 bins:  
30 bins both in $(0.12\sqrt{s} < E_\gamma< E_\chi^m)$ 
and in $(-0.99< \cos\theta_\gamma < 0.99)$, 
for the monophoton channel at CEPC. 
In each of the 900 bins, we compute the signal cross section in the 
dark matter models $\sigma^i_{\rm NP}$, 
and the SM cross section $\sigma^i_{\rm BG}$; 
we further sort the bins according to 
$r^i \equiv \sigma^i_{\rm NP}/\sigma^i_{\rm BG}$ 
in descending order so that the first bin has the 
largest signal to background ratio. 
We then take the first $n$ bins in the list to compute 
the 95\% CL upper bound and determine the 
$n$ value that leads to the best upper bound. 
We refer to this as the ``optimized cut'' hereafter. 
For the millicharged DM model in the CEPC $H$-mode, 
we find that typically $n \simeq 20$ (700) for the very low 
(very high) dark matter mass.

\begin{figure}[htbp]
\begin{centering}
\includegraphics[width=0.45\columnwidth]{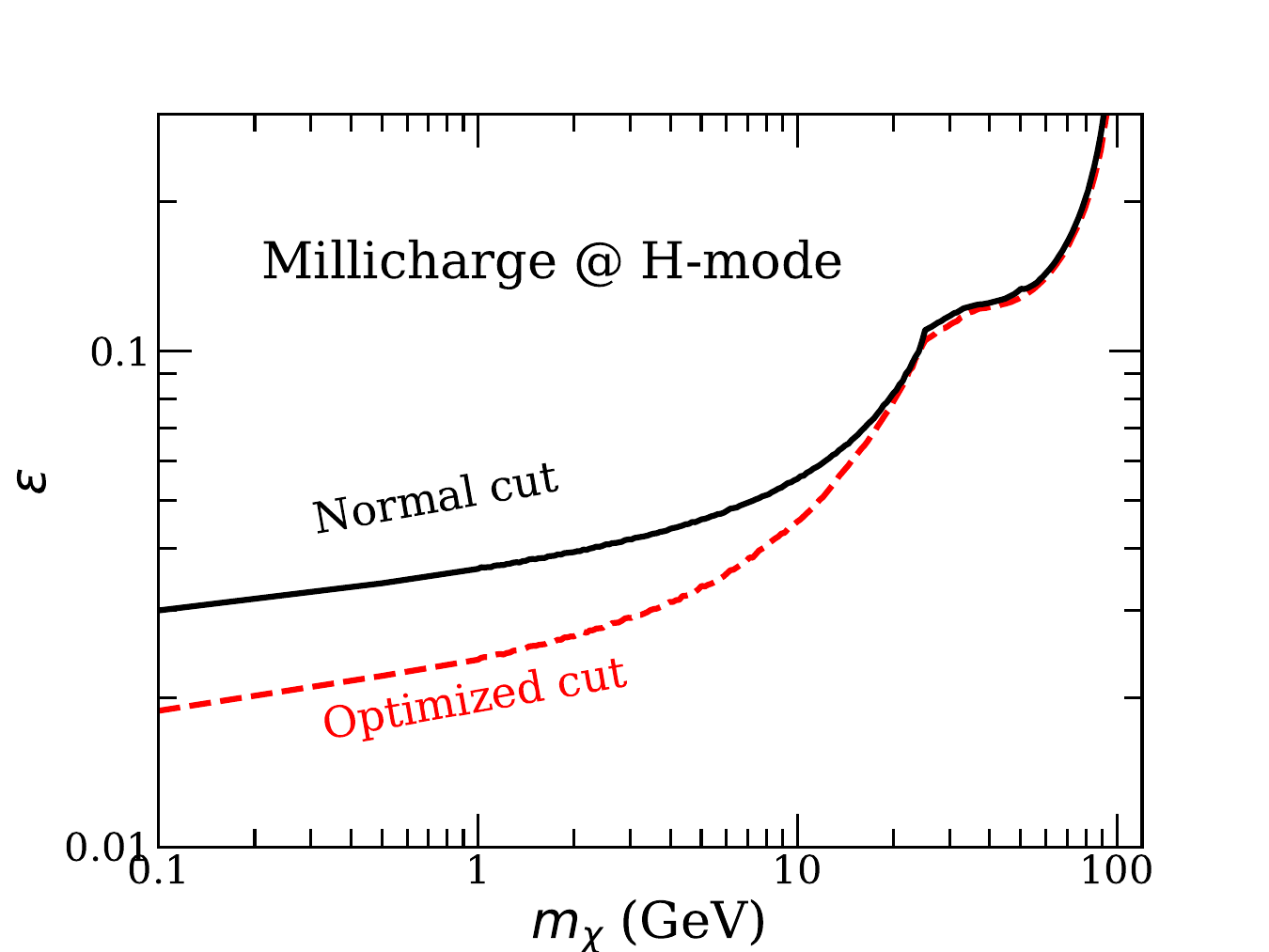}
\caption{Expected 95\% CL upper bound on millicharge 
analyzed with the ``optimizated'' cuts, and 
with the cuts used previously for millicharged DM models.}
\label{fig:optimsationresult}
\end{centering}
\end{figure}

Fig.\ (\ref{fig:optimsationresult}) shows the new limits 
on millicharge in the CEPC $H$-mode if we use the 
optimized cuts. The improvement is significant in 
the low DM mass regions. In particular, the expected 
upper bound on $\varepsilon$ goes from $\sim 0.03$ 
to $\sim 0.02$ for $m_\chi = 0.1\GeV$.


\section{Summary}
\label{sec:sum}

In this work, we investigate the capability of CEPC in probing 
millicharged DM models, $Z'$ portal DM models, 
and DM effective operators, 
by mainly using the monophoton channel. 
We propose a set of detector cuts which are found to 
efficiently suppress various background events and to 
improve the signal significance for the DM models considered.

For the millicharged DM models, the CEPC will probe 
the vast parameter space 
that is not constrained by 
previous collider experiments 
for ${\cal O}(1)-100$ GeV DM. 
The CEPC $Z$-mode with 16 ab$^{-1}$ 
has a better sensitivity than the other two running modes 
in the millicharged DM mass range $m_\chi \lesssim 40$ GeV, beyond which 
the $H$-mode dominates the reach.  
For $m_\chi \simeq 5$ (50) GeV, 
a new leading upper bound on millicharge 
$\varepsilon \lesssim 0.02$ (0.1)
is expected at CEPC. 
We further carry out an investigation on 
optimizing the detector cuts to constrain the 
millicharge; we found a significant improvement 
in the limits for low mass millicharged DM, 
by selecting only the high signal-to-background 
regions in the signature space.

For the $Z'$ portal DM model where $M_{Z'}=150$ GeV, 
we find that 
CEPC can explore the coupling down to $g^f\simeq 7\times10^{-4}$ 
for $g^\chi = 1$ in low DM mass region 
for both vector and axial vector couplings. 
An increased sensitivity on $g_V^f$ is 
observed for $m_\chi \simeq M_{Z'}/2$. 
We compare the constraining power of the monophoton channel 
with the $Z'$ visible decay channel and 
find that the visible channel is usually the better channel in 
parameter space where $g^f \gtrsim g^\chi$, 
and vice versa. 
For DM effective operators, 
the best constraints on the energy scale $\Lambda$ 
come from the CEPC $H$-mode in which 
$\Lambda$ up to $\simeq 1500\GeV$ in s-scalar, 
axial vector and vector effective operators 
and $\simeq 1100\GeV$ in t-scalar effective operators 
can be reached.

A complementary study 
on CEPC and dark matter direct detection limits from 
Xenon1T is carried out for $Z'$ portal and 
DM effective operators. 
For DM in the mass range 10-100 GeV, 
Xenon1T can cast a limit as good as CEPC or better 
assuming universal fermion couplings. 
CEPC provides better limits than the Xenon1T experiment, 
for GeV or sub-GeV dark matter particles, 
or in the case where DM only couples to electrons. 
We did not study 
the direct detection signal in Xenon1T for the 
millicharged dark matter particle, 
because it is likely to be absorbed in 
rock above the underground labs hosting 
dark matter direct detection experiments, for 
the millicharge of interest at CEPC searches.


\acknowledgments

We thank Zihao Xu for discussions. 
The work is supported in part  
by the National Natural Science Foundation of China under Grant Nos.\ 
11775109, U1738134 and 11805001, 
and by the National Recruitment Program for Young Professionals.


\end{document}